\documentclass[]{spie}  

\usepackage{amsmath,amsfonts,amssymb}
\usepackage{txfonts}  
\usepackage{graphicx}
\usepackage{xcolor,colortbl}
\usepackage[colorlinks=true, allcolors=blue]{hyperref}

\title{Upgrading the high contrast imaging facility SPHERE: science drivers and instrument choices}

\author[a]{A. Boccaletti} 
\author[b]{G. Chauvin} 
\author[c]{F. Wildi} 
\author[d]{J. Milli} 
\author[d]{E. Stadler} 
\author[e]{E. Diolaiti} 
\author[f]{R. Gratton} 
\author[a]{F. Vidal} 
\author[g]{M. Loupias} 
\author[g]{M. Langlois} 
\author[h]{F. Cantalloube} 
\author[b]{M. N'Diaye} 
\author[a]{D. Gratadour} 
\author[a]{F. Ferreira} 
\author[g]{M. Tallon} 
\author[a]{J. Mazoyer} 
\author[c]{D. Segransan} 
\author[d]{D. Mouillet} 
\author[h]{J.-L. Beuzit} 
\author[d]{M. Bonnefoy} 
\author[a]{R. Galicher} 
\author[h]{A. Vigan} 
\author[i]{I. Snellen} 
\author[j]{M. Feldt} 
\author[f]{S. Desidera} 
\author[b]{S. Rousseau} 
\author[f]{A. Baruffolo}
\author[a]{C. Goulas} %
\author[a]{P. Baudoz} 
\author[g]{C. Bechet} 
\author[b]{M. Benisty} 
\author[k]{A. Bianco}
\author[b]{B. Carry} 
\author[l]{E. Cascone}
\author[a]{B. Charnay} 
\author[h]{E. Choquet} 
\author[m]{V. Christiaens} 
\author[e]{F. Cortecchia}
\author[l]{V. de Caprio}
\author[e]{A. De Rosa}
\author[d]{C. Desgrange} 
\author[f]{V. D'Orazi}
\author[d]{S. Douté} 
\author[k]{M. Frangiamore}
\author[a]{E. Gendron} 
\author[i]{C. Ginski} 
\author[a]{E. Huby} 
\author[i]{C. Keller} 
\author[n]{C. Kulcs\'ar} 
\author[i]{R. Landman} 
\author[b]{S. Lagarde} 
\author[b]{E. Lagadec} 
\author[a]{A.-M. Lagrange} 
\author[e]{M. Lombini}
\author[o]{M. Kasper} 
\author[d]{F. Ménard} 
\author[d]{Y. Magnard} 
\author[e]{G. Malaguti}
\author[d]{D. Maurel} 
\author[f]{D. Mesa}
\author[e]{G. Morgante}
\author[p]{E. Pantin} 
\author[p]{T. Pichon} 
\author[q]{A. Potier} 
\author[d]{P. Rabou} 
\author[d]{S. Rochat} 
\author[e]{L. Terenzi}
\author[g]{E. Thiébaut} 
\author[g]{I. Tallon-Bosc} 
\author[n]{H.-F. Raynaud} 
\author[a]{D. Rouan} 
\author[a]{A. Sevin} 
\author[e]{F. Schiavone}
\author[e]{L. Schreiber}
\author[k]{A. Zanutta}
\affil[a]{LESIA, Observatoire de Paris, Université PSL, CNRS, Sorbonne Université, Université de Paris Cité, 5 place Jules Janssen, 92195 Meudon, France}
\affil[b]{Université Côte d'Azur, Observatoire de la Côte d'Azur, CNRS, Laboratoire Lagrange, France}
\affil[c]{Départment d’astronomie de l’Université de Genève, 51 ch. des Maillettes Sauverny, 1290 Versoix, Switzerland}
\affil[d]{Univ. Grenoble Alpes, CNRS, IPAG, 38000 Grenoble, France}
\affil[e]{INAF Osservatorio di Astrofisica e Scienza dello Spazio di Bologna, Via P. Gobetti 93/3, 40129, Bologna, Italy}
\affil[f]{INAF Osservatorio Astronomico di Padova, Vicolo dell’Osservatorio 5, 35122, Padova, Italy}
\affil[g]{Univ Lyon, Univ Lyon1, Ens de Lyon, CNRS, Centre de Recherche Astrophysique de Lyon UMR5574, F-69230, Saint-Genis-Laval, France}
\affil[h]{Aix Marseille Université, CNRS, LAM, UMR 7326, 13388, Marseille, France}
\affil[i]{Leiden Observatory, Leiden University, PO Box 9513, 2300 RA Leiden, The Netherlands}
\affil[j]{Max Planck Institute for Astronomy, K\"onigstuhl 17, D-69117 Heidelberg, Germany}
\affil[k]{INAF Osservatorio Astronomico di Brera, Via Brera 28, 20121, Milano, Italy}
\affil[l]{INAF Osservatorio Astronomico di Capodimonte, Salita Moiariello 16, 80131 Napoli, Italy}
\affil[m]{Space sciences, Technologies and Astrophysics Research (STAR) Institute, University of Liège, 19C allée du Six Août, 4000 Liège, Belgium}
\affil[n]{Université Paris-Saclay, Institut d'Optique Graduate School, CNRS, Laboratoire Charles Fabry, Palaiseau, France}
\affil[o]{European Southern Observatory, Karl-Schwarzschild-Str. 2, 85748, Garching, Germany}
\affil[p]{Université Paris-Saclay, Université Paris Cité, CEA, CNRS, AIM, 91191, Gif-sur-Yvette, France}
\affil[q]{Jet Propulsion Laboratory, California Institute of Technology, 4800 Oak Grove Drive, Pasadena, CA 91109}

\begin{document} 
\maketitle

\begin{abstract}

SPHERE+ is a proposed upgrade of the SPHERE instrument at the VLT, which is intended to boost the current performances of detection and characterization for exoplanets and disks. SPHERE+ will also serve as a demonstrator for the future planet finder (PCS) of the European ELT. The main science drivers for SPHERE+ are 1/ to access the bulk of the young giant planet population down to the snow line ($3-10$ au), to bridge the gap with complementary techniques (radial velocity, astrometry); 2/ to observe fainter and redder targets in the youngest ($1-10$\,Myr) associations compared to those observed with SPHERE to directly study the formation of giant planets in their birth environment; 3/ to improve the level of characterization of exoplanetary atmospheres by increasing the spectral resolution in order to break degeneracies in giant planet atmosphere models. Achieving these objectives requires to increase the bandwidth of the xAO system (from $\sim$1 to 3\,kHz) as well as the sensitivity in the infrared (2 to 3\,mag). These features will be brought by a second stage AO system optimized in the infrared with a pyramid wavefront sensor. As a new science instrument, a medium resolution integral field spectrograph will provide a spectral resolution from 1000 to 5000 in the J and H bands. This paper gives an overview of the science drivers, requirements and key instrumental trade-off that were done for SPHERE+ to reach the final selected baseline concept. 
\end{abstract}


\section{INTRODUCTION}
\label{sec:intro} 

Understanding how and when planets form is a central question in the study of exoplanets. The 5000+ known exoplanets are all valuable to provide insight to this question either from an individual point of view, or in a statistical sense. Among the techniques used to detect and study exoplanets, direct imaging occupies a special place since it is restrained to the observation of young stars for contrast reasons, which in turn also has the ability to provide more straightforward constraints on planet formation than with other methods. The case of the system PDS\,70 where two giant planets are caught at formation is a good illustration of this advantage . Direct imaging is also one of the two methods able to directly perform spectral characterization of exoplanet atmospheres. This has been done at relatively low spectral resolution so far, but the field is now moving toward medium to high spectral resolution.

In the context of direct imaging of exoplanets, SPHERE, the Spectro-Polarimetic High contrast imager for Exoplanets REsearch \cite{Beuzit2019} has been instrumental to collect very detailed information on exoplanetary systems in a broad sense, that is exoplanets and their environments.  SPHERE provided new insights on a variety of brown dwarf companions \cite{Cheetham2018, Maire2016, Delorme2017}, as well as known planetary mass companions \cite{Zurlo2016, Bonnefoy2016, Chauvin2018, Samland2017}, which paved the way towards the first detection of a cloudy massive Jupiter, HIP\,65426 b \cite{Chauvin2017}. Then, the first unambiguous discovery of a young planet, PDS\,70 b, in a formation stage has followed \cite{Keppler2018, Muller2018}. Near IR photometry and low resolution spectroscopy of these young giant planets has allowed to explore the physical processes at play in their atmospheres, confirming low surface gravities, and the presence of thick dusty clouds. With the exquisite stability of SPHERE it was also possible to study these systems dynamically, the follow up of $\beta$ Pic b, the HR\,8799 system, as well as 51 Eri b being good examples \cite{Lagrange2019, Maire2019}. 
One of the greatest achievements of SPHERE has been to reveal the fine structures within imaged circumstellar disks, yielding unprecedented information regarding the structure of planetary systems during or right after the planetary formation phase. Large field of view, and polarimetric capabilities has been decisive to resolve a wealth of structures in protoplanetary disks: large spiral arms \cite{Benisty2015}, multiple rings or gaps \cite{vanBoekel2017}, variable shadows \cite{Stolker2016}, and central cavities \cite{Pohl2017, Ligi2018}, and similarly for debris disks: sharp Kuiper-like belts \cite{Lagrange2016, Milli2017}, multiple belts in gas-rich debris disks \cite{Perrot2016, Feldt2017}, moving structures \cite{Boccaletti2015}. The vast majority of these structures could be explained by the presence of planets or by massive collisions. Disk science is closely connected to exoplanet science, as reinforced by the results described above and especially the case of PDS 70 b, where a planet was discovered right inside the cavity of a gas-rich disk.

\begin{figure*}[t]
\begin{center}
\includegraphics[trim=1cm 0 1cm 0, height=6.5cm]{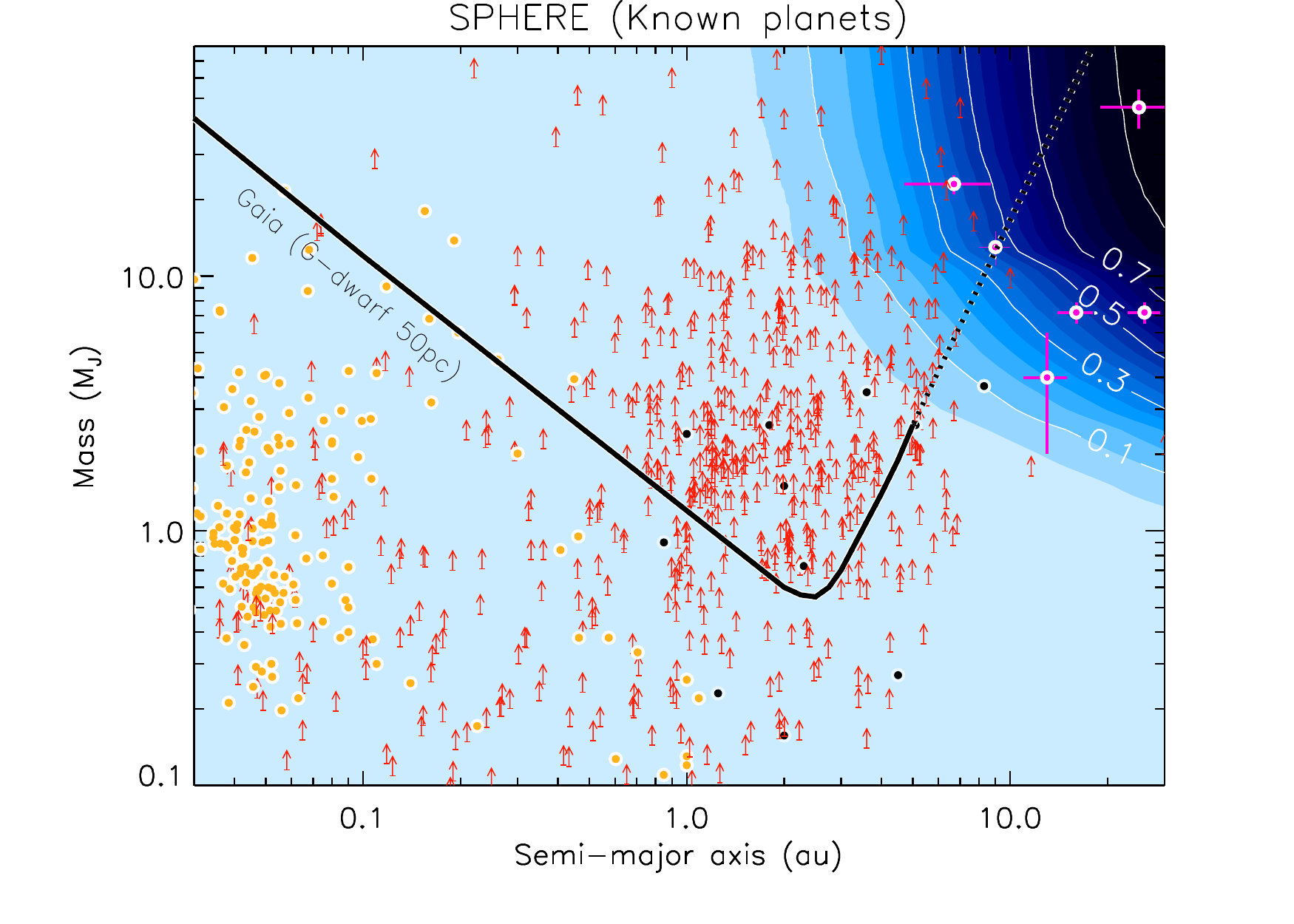}
\includegraphics[trim=1cm 0 1cm 0, height=6.5cm]{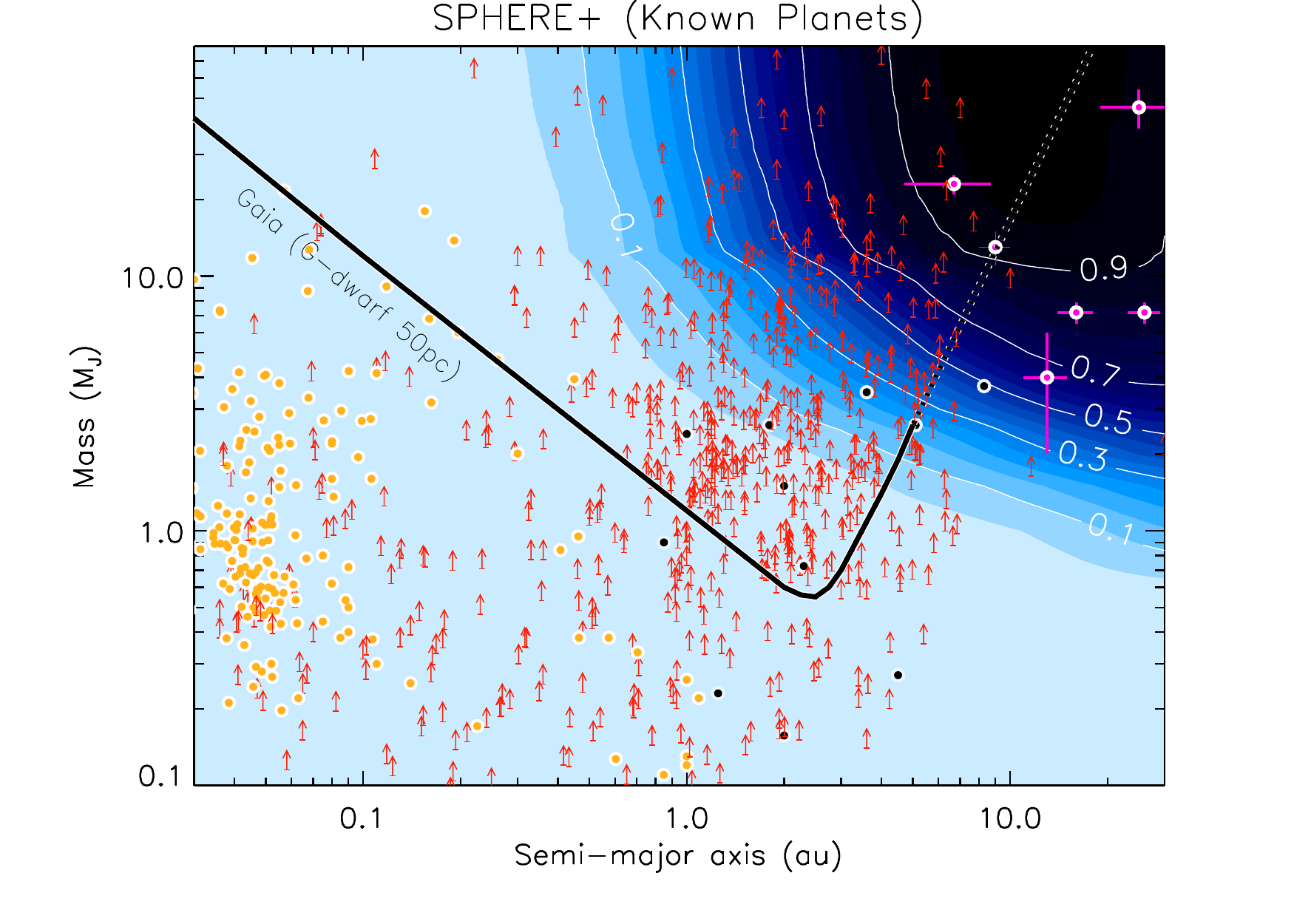}\\
\caption{SPHERE (\textit{Left}) and expected SPHERE+ (\textit{Right}) detection probabilities from the SHINE sample of young ($\sim 100$\,Myr), nearby ($\sim 50$\,pc) stars compared to the current population of exoplanets detected with all techniques: transit (\textit{yellow} dot), radial velocity (\textit{red} arrow), $\mu$-lensing (\textit{black} dot), and direct imaging (\textit{pink} dot). The \textit{Gaia} detection limits that will be available with the Data Release 4 for a solar-type star at 50\,pc are overlaid. }
\label{fig:bridge}
\end{center}
\end{figure*}

One important feature of the guaranteed time observations is the SpHere INfrared survey of Exoplanets (SHINE) which observed $\sim$600 stars. The first analysis of 150 targets demonstrates that giant planets are relatively rare beyond 10\,au \cite{Vigan2021} (Fig. \ref{fig:bridge}). On the contrary, radial velocity surveys \cite{Fernandes2019}, suggest the bulk of the population being located at $3-10$\,au. However, the current performance of SPHERE \cite{Langlois2021} in this physical separation range will not allow us to fully bridge the gap in planet separation in order to image planets detected via complementary techniques (transit, radial velocity, $\mu$-lensing and astrometry).  Thus, further improvements in contrast at short angular separation will be necessary to offer a global vision of planet occurrence at all separations.

\begin{figure*}[t]
\begin{center}
\includegraphics[trim = 0cm 0cm 0cm 0cm,, height=6.5cm]{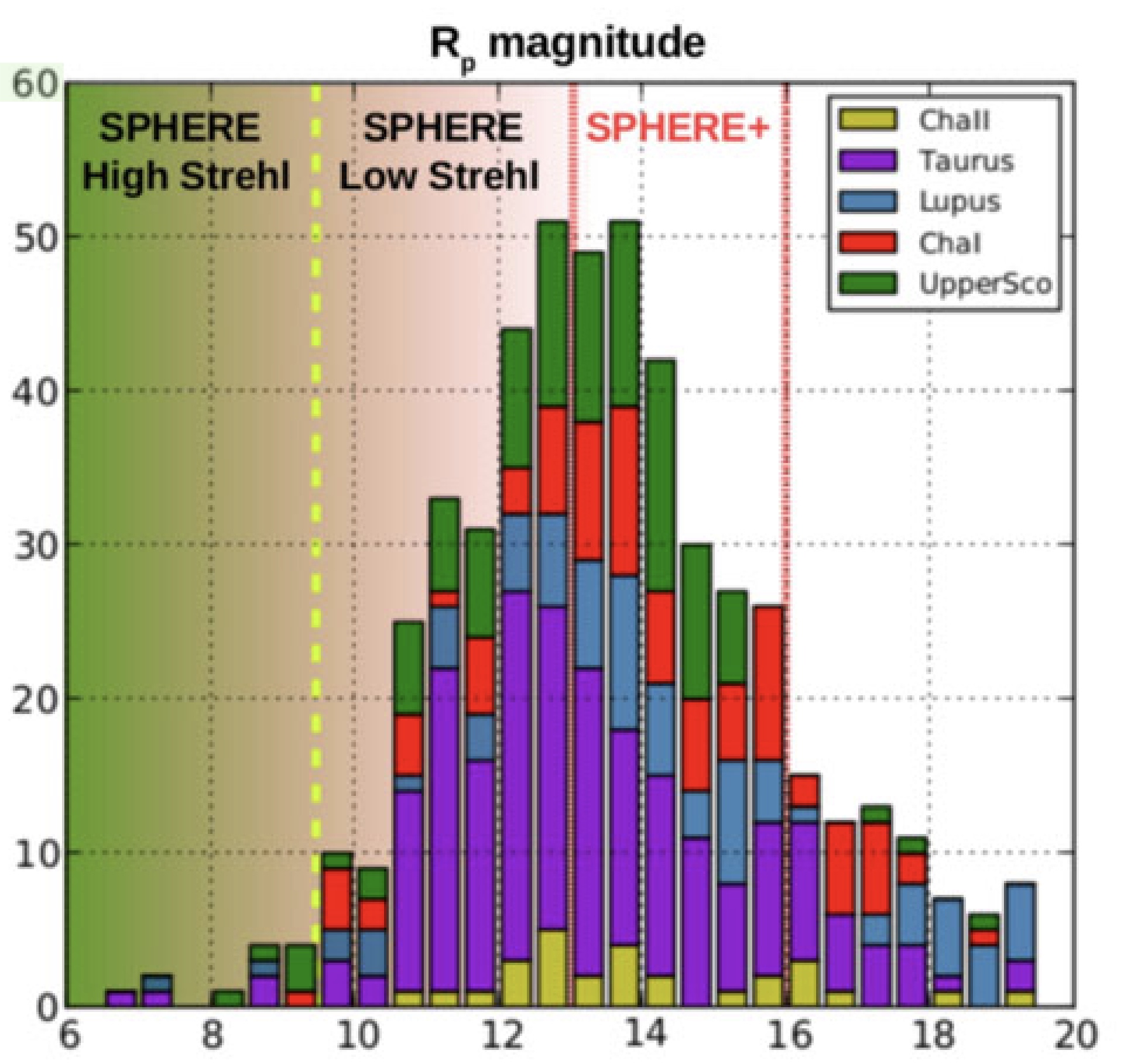}
\includegraphics[trim=0cm 0cm 0cm 0cm, height=6.5cm]{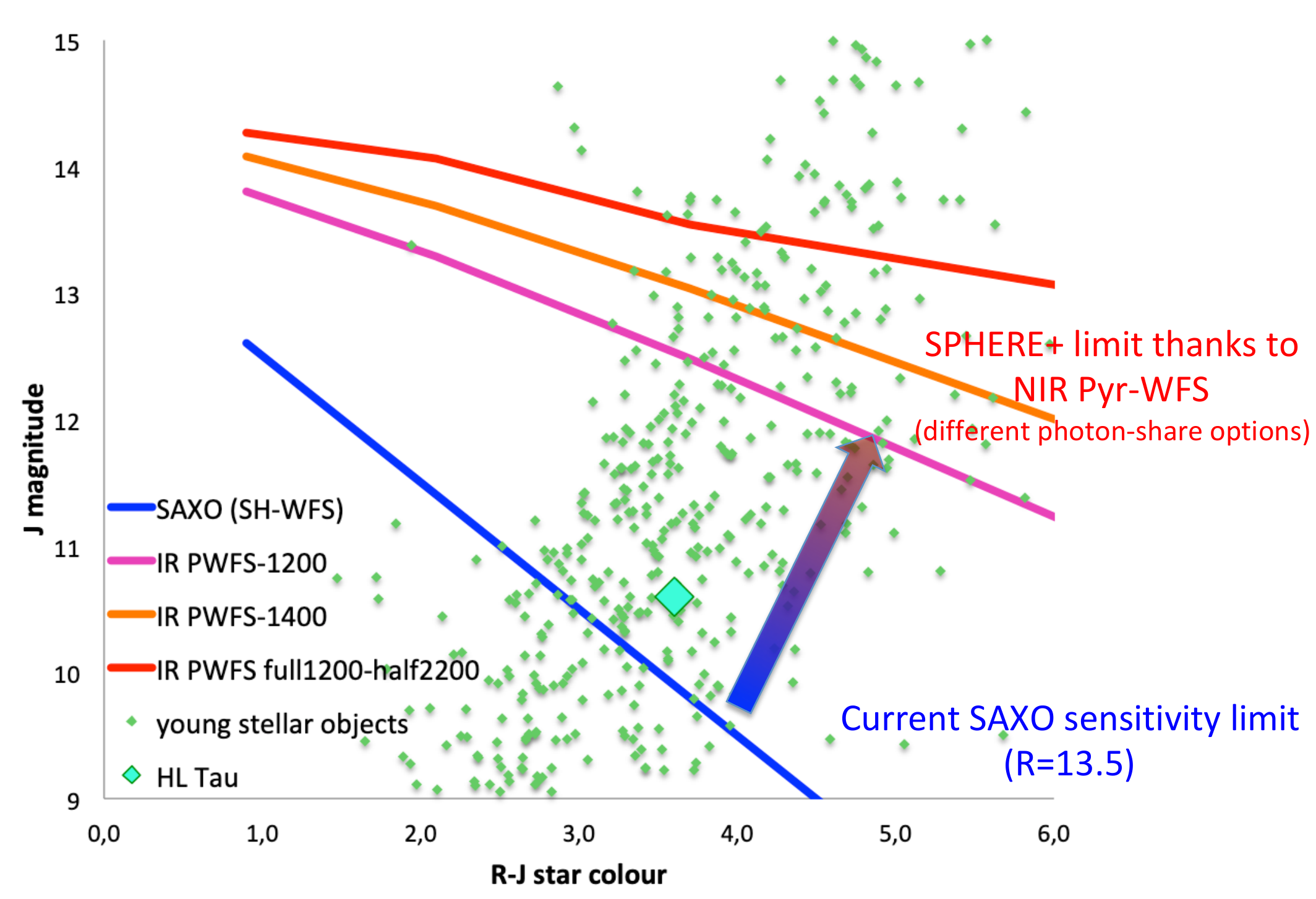}\\
\caption{Left: histogram of target's R magnitude  for several young star associations, indicating the performance of SPHERE and those expected with SPHERE+. Right: Magnitude color diagram of actual targets in star forming regions observable with ALMA (including the famous HL Tau) overlaid with the AO sensitivity of SAXO (blue line) and the sensitivity expected with an IR Pyramid WFS for several assumptions for the spectral range.}
\label{fig:redtargets}
\end{center}
\end{figure*}

Another limitation comes from a low sensitivity (red/faint targets) to planet-forming disks observed in the submillimeter (Fig. \ref{fig:redtargets}), while the combined analysis of ALMA and SPHERE data has proven promising to diagnose planet formation conditions \cite{Boccaletti2020a, Benisty2021}. Finally, the analyzing of planet’s atmospheric properties (chemical compounds, thermal structures, clouds) is strongly limited by the low spectral resolution (R=50). 

The SPHERE+ project has been proposed to circumvent the main limitations of SPHERE in terms of contrast at short angular separations, sensitivity to faint and red targets, and spectral resolution for atmosphere characterization. For that purpose the main new components that are considered for a SPHERE upgrade are: 1/ a second stage AO system, SAXO+, to achieve faster correction rate with an IR wavefront sensor, and 2/ a new integral field spectrograph, MEDRES, to achieve a spectral resolution of about 1000 at near IR wavelengths.


\section{Science drivers and technical requirements}

The main motivation for SPHERE+ relies on three key scientific requirements that are currently driving this project as well as our proposed instrumental concept. They can be summarized as follows : 

\begin{itemize}
    \item \texttt{sci.req.1} - {\bf Access the bulk of the young giant planet population down to the snow line} ($3-10$ au), in order to bridge the gap with complementary techniques (radial velocity, astrometry), taking advantage of the synergy with \textit{Gaia}, and to explore for the first time the complete demographics of young giant planets at all separations in order to constrain their formation and evolution mechanisms (Fig. \ref{fig:bridge}).
    
    \item  \texttt{sci.req.2} - {\bf Observe a large number of fainter (lower mass) stars} in the youngest ($1-10$\,Myr) associations (Lupus, Taurus, Chamaeleontis, Scorpius-Centaurus...), to directly study the formation of giant planets in their birth environment, building on the synergy with ALMA to characterize the architectures and properties of young protoplanetary disks, and how they relate to the population of planets observed around more evolved stars (Fig. \ref{fig:redtargets}).
    
    \item  \texttt{sci.req.3} -  {\bf Improve the level of characterization of exoplanetary atmospheres} by increasing the spectral resolution in order to break degeneracies in giant planet atmosphere models and to measure abundances and other physical parameters, such as the radial and rotational velocities. 
\end{itemize}

Overall, the SPHERE+ top level requirements connected to the proposed science cases can be summarized by going {\bf closer}, {\bf deeper}, and {\bf fainter}. As we understand very well the limitations of SPHERE, the science requirements can be linked directly to the following instrumental requirements:
\begin{itemize}
    \item \texttt{tech.req.1} - Deeper/closer: {\bf increase the bandwidth of the xAO system} (typically 3kHz instead of 1kHz) and improve the correction of non-common path aberrations and coronagraphic rejection.
    \item \texttt{tech.req.2} - Fainter: include a {\bf more sensitive wavefront sensor} to gain 2-3 magnitudes for red stars.
    \item  \texttt{tech.req.3} - Enhanced characterization: {\bf develop spectroscopic facilities with significantly higher spectral resolution} compared to the current SPHERE Integral Field Spectrograph (IFS). In this respect, both medium ($R_\lambda=1\,000-5\,000$) and high ($R_\lambda=50\,000-100\,000$) resolutions are extremely valuable for the characterization of planetary atmospheres. 
\end{itemize}

\section{SAXO+}

The current AO system of SPHERE, SAXO\cite{Fusco2014}, is made of a spatially filtered Shack-Hartmann wavefront sensor with 40x40 sub-apertures, combined to a low-noise EMCCD detector ($<0.1$\,e-) operating at visible wavelengths, and a high order 41x41 deformable mirror by Cilas (HODM). The Real Time Computer, SPARTA, is running the loop at a maximum frequency of 1.38 kHz  on bright stars. SAXO has demonstrated Strehl ratio in the 90\% range in the H band, up to R magnitude of 10 \cite{Sauvage2016}, and even reaching moderate Strehl ratio up to R=14. The main constraint for an upgrade is to preserve all the functionalities and the performances of SPHERE both in the visible and near IR arms, so that any improvement should be considered as additional modes with the capability to switch back to the initial configuration. SAXO+ is designed to operate by default with IRDIS and IFS (in simultaneous observations as in the current configuration), but also with MEDRES and HIRiSE \cite{Vigan2018,Otten2021,Vigan2022spie}.

The science requirements for SPHERE+ are driving the characteristics of SAXO+. To gain contrast at short separations around bright stars, the 2nd stage should run faster, in a typical range of 2 to 3 kHz. The exact frequency is to be optimized with AO simulations. To allow to observe targets that have J magnitudes of 10 to 13 but with R-J of 3 to 5, SAXO+ should implement an IR wavefront sensor (Fig. \ref{fig:redtargets}). We choose a Pyramid Wavefront sensor for sensitivity reasons. As we intend to focus primarily on a narrower field of view than SPHERE, the deformable mirror can be of lower order compared to the 1st stage, so we are considering a 24x24 or 32x32 MEMS DM. 

A key question for SAXO+ is the level of interaction between the two AO stages for optimal performances. Our baseline, which is essentially constrained by \texttt{sci.req.2}, is to have an "integrated solution" where one single new RTC drives the two loops. That means the wavefront measurements are obtained from the visible Shack-Hartmann and the IR Pyramid, while the correction is performed with both the HODM and the new faster DM. The RTC is based on the COSMIC concept \cite{Ferreira2018} while the current RTC, SPARTA, would be bypassed. An alternative is to control the full system with just the IR WFS, if it proves to be efficient enough. This derivative case will also be studied with simulations.
In that integrated version, SAXO+ is more like an extension of the current AO system than an additional component. The second solution is a "standalone", with no real time interactions between the two loops, the first one still being controlled by SPARTA, and the second stage by COSMIC. We anticipate that this solution could be acceptable for bright stars (\texttt{sci.req.1}) but can be problematic to handle the science case of faint and red targets as long as the first stage will certainly makes the correction worse. For this reason the "standalone" solution is only considered as a backup plan.

In terms of photon sharing, we identified that two or three dichroic beamsplitters will be needed depending on the redness of the targets to optimize the photon flux on the wavefront sensor. The two first dichroic beamsplitters are optimized for the spectral bands Y ($0.95-1.2\,\muup$m) and YJ ($0.95-1.4\,\muup$m), while a third one can extend to the H band with 50\% of transmission ($0.95-1.4\,\muup$m + half of $1.4-1.8\,\muup$m).

\begin{figure*}[t]
\begin{center}
\includegraphics[trim = 0cm 0cm 0cm 0cm,, height=6.cm]{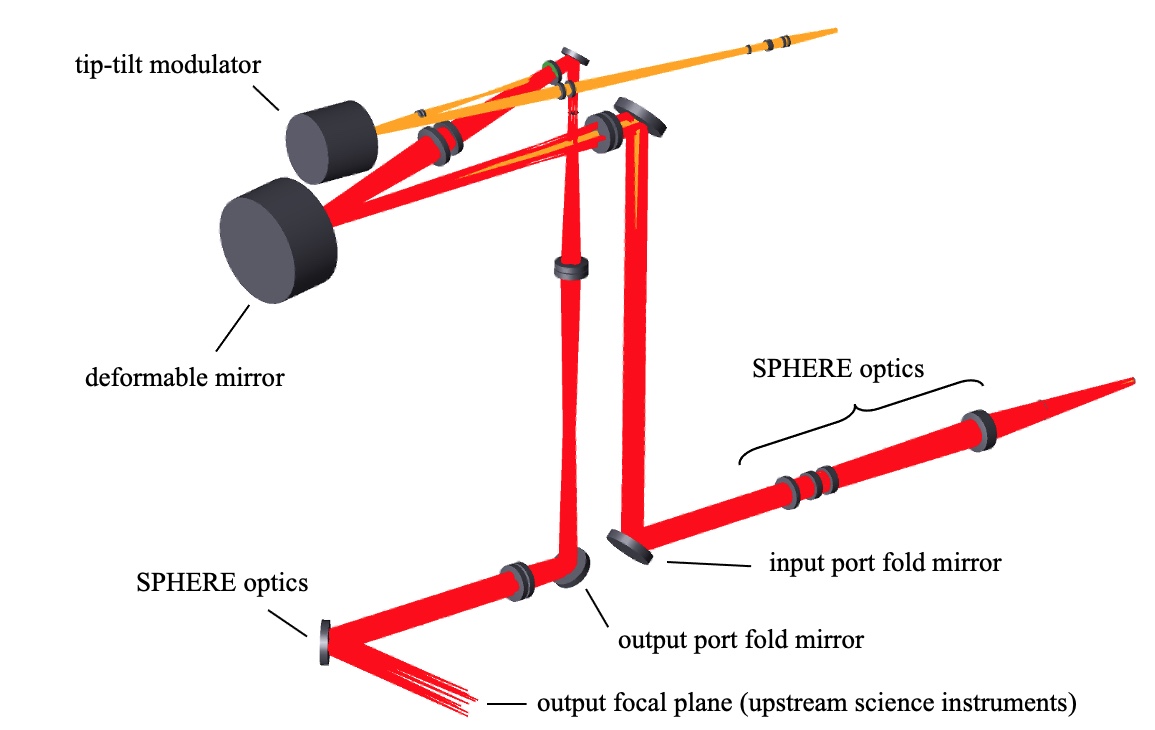}
\includegraphics[trim=0cm 0cm 0cm 0cm, height=6.cm]{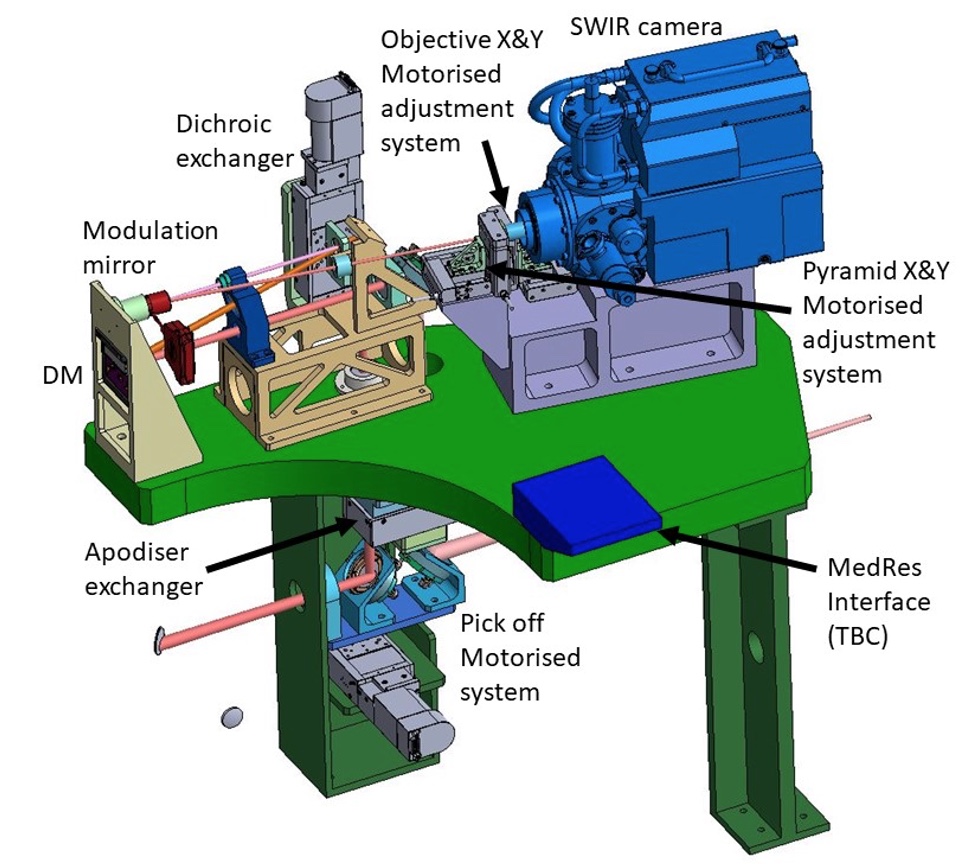}\\
\caption{Optical (left) and mechanical (right) design of SAXO+. In the left panel, the red beam stands for the science path, while the orange one is the wavefront sensor path. The right panel shows the implementation of the optical elements in the mechanical frame,  inside the green bench which resides on top on the SPHERE common path. The drawing indicates the optical mechanical supports, the motors for moving parts and the detector (blue camera).}
\label{fig:saxoplus}
\end{center}
\end{figure*}

Given the aforementioned constraints, we designed SAXO+ has a separate compact module which fits downstream the 1st stage, in between the Atmospheric Dispersion Corrector and the coronagraphic mask/Differential Tip Tilt Sensor. The current status of the SAXO+ opto-mechanical design is fully detailed in Stadler et al. (this proceeding, paper 12185-165). Here we summarize the main features of this sub-system. A pick off mirror is installed to send the IR beam upwards, 40\,cm above the SPHERE optical axis. The two active components, that is the DM and the modulator for the pyramid are both located close to the enclosure for maintenance purpose. The baseline DM is 10mm in diameter, but we can accommodate the Boston Micro Machine kilo-DM which is 13mm in diameter. All components of the science channel are in a single vertical plane with small incidence angles to preserve polarimetric capabilities with IRDIS. The wavefront sensing channel after the dichoic beamsplitter is slightly tilted with respect to this plane. The focal ratio at the pyramid is at least F/20 (goal F/40). The output beam is at F/40 for compliance with the other optics in SPHERE, in particular the coronagraphic mask in the common path and the stop in IRDIS. The current coronagraph's apodizers located on a wheel inside the ADC would prevent wavefront sensing with SAXO+, so a dedicated pupil plane is needed inside SAXO+ right upstream of the corongraphic mask. As a baseline we will implement the very same apodizers \cite{Carbillet2011} for consistency with former observations but we will also explore new apodizer designs to improve the contrast at close angular separations as it was identified as a limitation \cite{Vigan2019}. Figure \ref{fig:saxoplus} shows the optical and mechanical layout of SAXO+.

\begin{figure*}[t]
\begin{center}
\includegraphics[trim = 0cm 0cm 0cm 0cm, height=5cm]{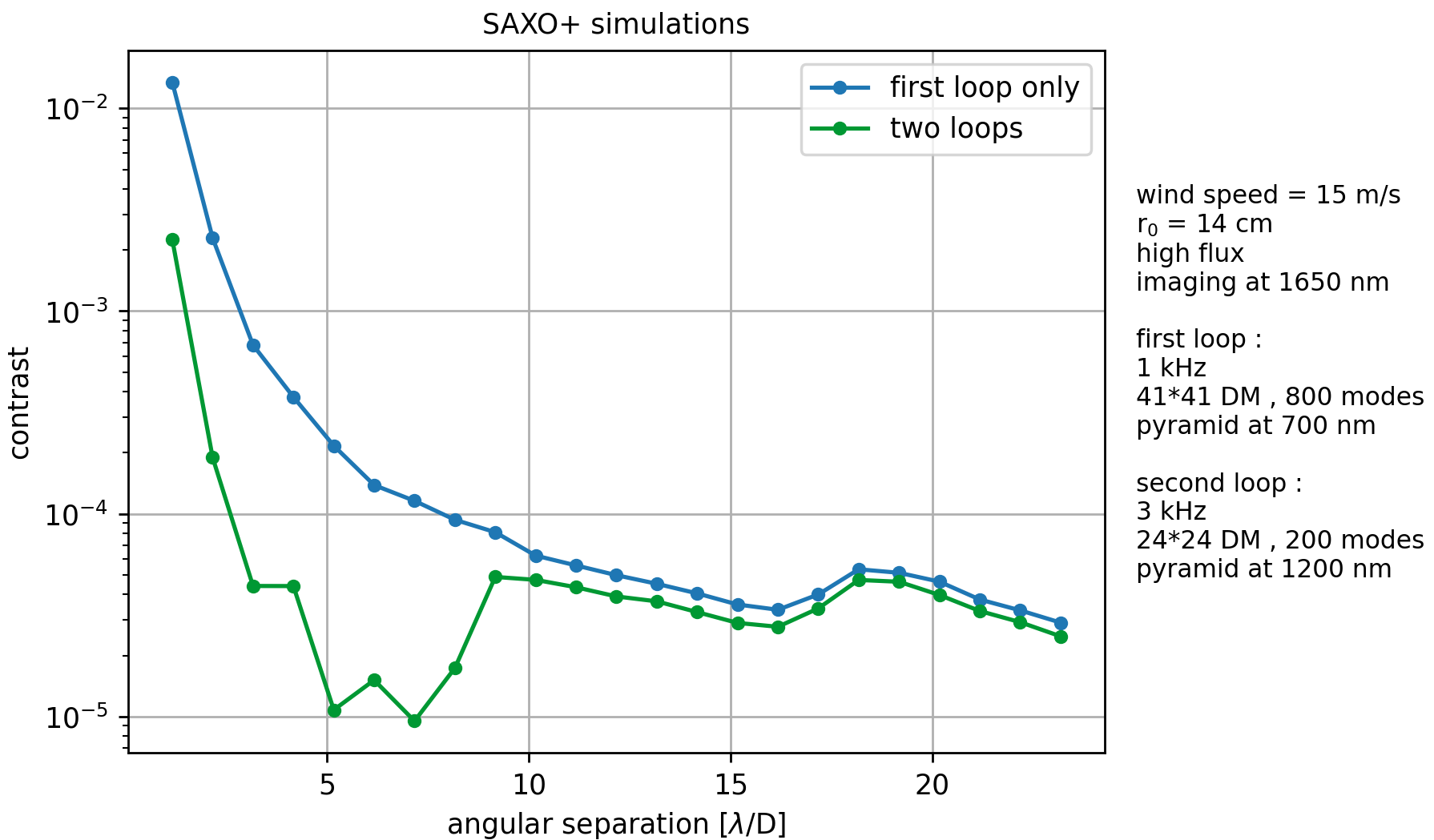}
\includegraphics[trim = 0cm 0cm 0cm 0cm, height=5cm]{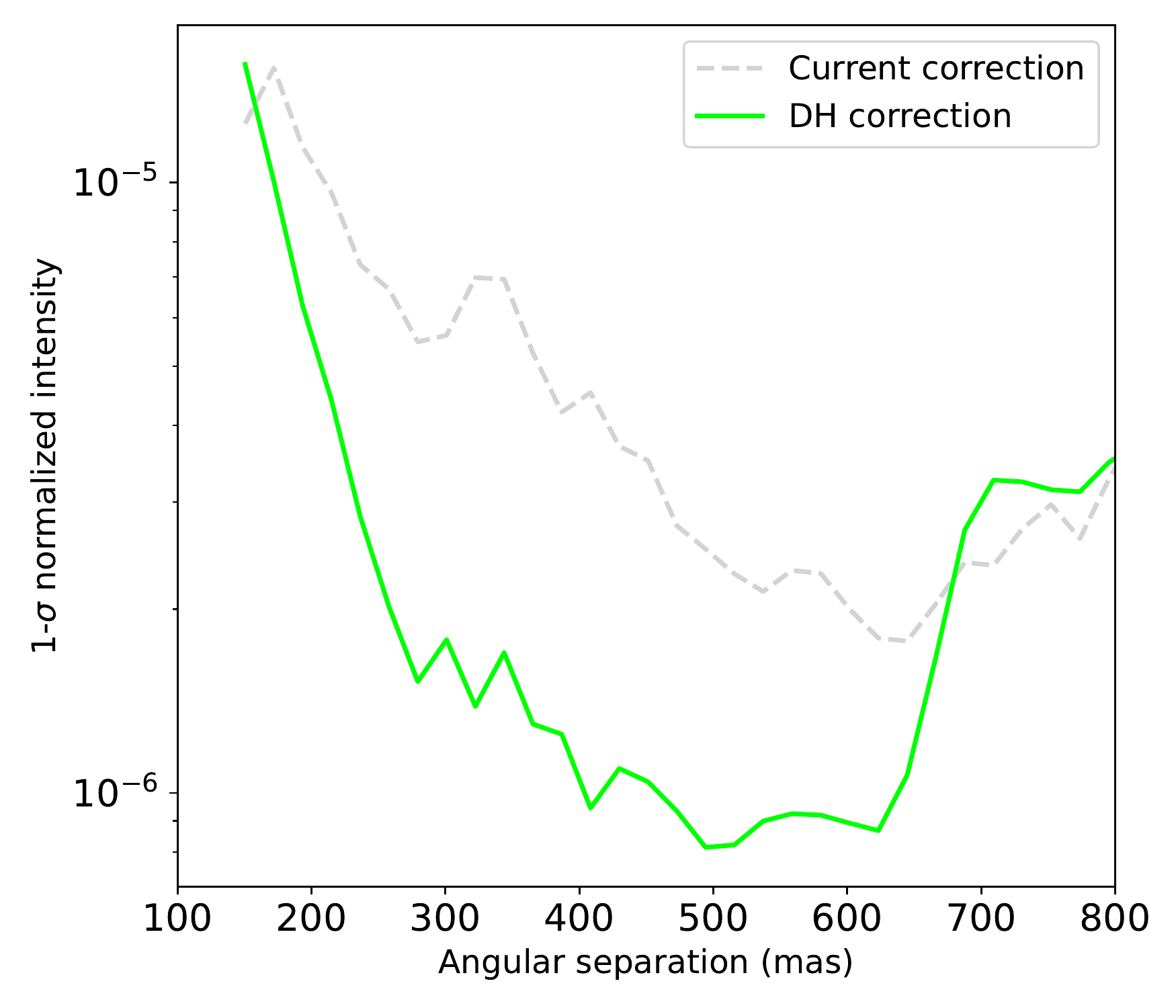}
\caption{Left: simulated 5$\sigma$ contrast with SAXO (blue) and SAXO+ (green) for a typical set of atmospheric conditions. Right: 1$\sigma$ contrast obtained with the dark hole technique on sky with SPHERE (green line) as compared to the nominal case (grey line).}
\label{fig:simulsaxoplus}
\end{center}
\end{figure*}

Preliminary assessment of performances with two AO loops are presented in Vidal et al. (this proceeding, paper 12185-164). In this first estimation we assumed typical turbulence conditions (wind speed 15\,m/s, r$_0=14$\,cm, bright target), and a first AO loop simulated with COMPASS using the characteristics of SAXO (frequency 1\,kHz, 41$\times$41 actuators DM, 800 modes corrected, waveferont sensing measurement at 700nm), and a first guess of the SAXO+ characteristics (frequency 3\,kHz, 24$\times$24 actuators DM, 200 modes corrected, waveferont sensing measurement at 1200nm). Beside, we simulated a "perfect" coronagraph, which subtracts a perfect PSF normalized to the Strehl ratio. The result shown in Fig. \ref{fig:simulsaxoplus} (left) is promising as we can expect a gain as large as an order of magnitude on the raw contrast at a separation of 5 $\lambda$/D. This has to be confirmed with more accurate simulations taking into account temporal effects between the two AO loops, low-wind effect and non common path aberrations, as well as real coronagraphs.

In addition to a second faster AO loop we foresee the implementation of a third slower loop using the scientific image as a sensor with the ability to produce a dark-hole in the corrected AO region. Potier et al., 2020 \cite{Potier2020} demonstrated the capability of the technique on the SPHERE internal source, which combines pair-wise probing that provides a modulation of the speckle field with the DM to estimate the post-AO beam splitter wavefront aberrations and Electric Field Conjugation to calculate the correction. Both phase and amplitude can be compensated in half of the corrected area. More recently, on-sky experiment has been proven successful, achieving contrasts as large as $10^{-6}$ or better (1$\sigma$ contrast) at about 500\,mas (Fig. \ref{fig:simulsaxoplus}, right), which represents a factor of 5 in improvement of the raw contrast (Potier et al. 2022, submitted, and this proceeding paper 12185-236). As a comparison, the median contrast of the SPHERE survey \cite{Langlois2021} is somewhat equivalent but requiring about 1 hour of observation with angular differential imaging. The ability to obtain similar performances in just 1 minute is extremely promising for the future of the SPHERE observations. Implemented in SPHERE+, the dark-hole method will be fully integrated in the RTC with potentially even more stable and robust contrasts. Further improvement is achievable using coherent differential imaging as a postprocessing technique to reach a few $10^{-6}$ contrast level in the other half of the AO corrected region.

\section{MEDRES}

IFS, the integral field spectrograph of SPHERE provides a broad spectral range with two spectral resolutions: the YJ mode at R=50, and the YJH mode at R=30, in a 1.7$''$x1.7$''$ field of view. IFS data reduction takes advantage of Angular and Spectral Differential Imaging to reach very high contrasts. These methods exploit the angular and spectral diversity between the speckles and the image of a real companion; they are extremely powerful at apparent separation of a few $\lambda/D$ from the star, but are inefficient at very short separation. Recently another method has been proposed to improve the detection of exoplanet atmospheres which is based on cross correlating the spaxels in IFS data with molecular templates estimated from exoplanet atmosphere models, also referred as "molecular mapping". This technique is extremely powerful to disentangle the signature of a planet amongst the halo of speckles which have different spectral behavior, and it has a lower dependency with the angular separation compared to ADI and ASDI. It has been successful to study the atmosphere of $\beta$ Pic b \cite{Hoeijmakers2018} and HIP\,65426 \cite{Petrus2020} at a spectral resolution of about 4000 with SINFONI. 

\begin{figure*}[t]
\begin{center}
\includegraphics[trim = 0cm 0cm 0cm 0cm, height=5cm]{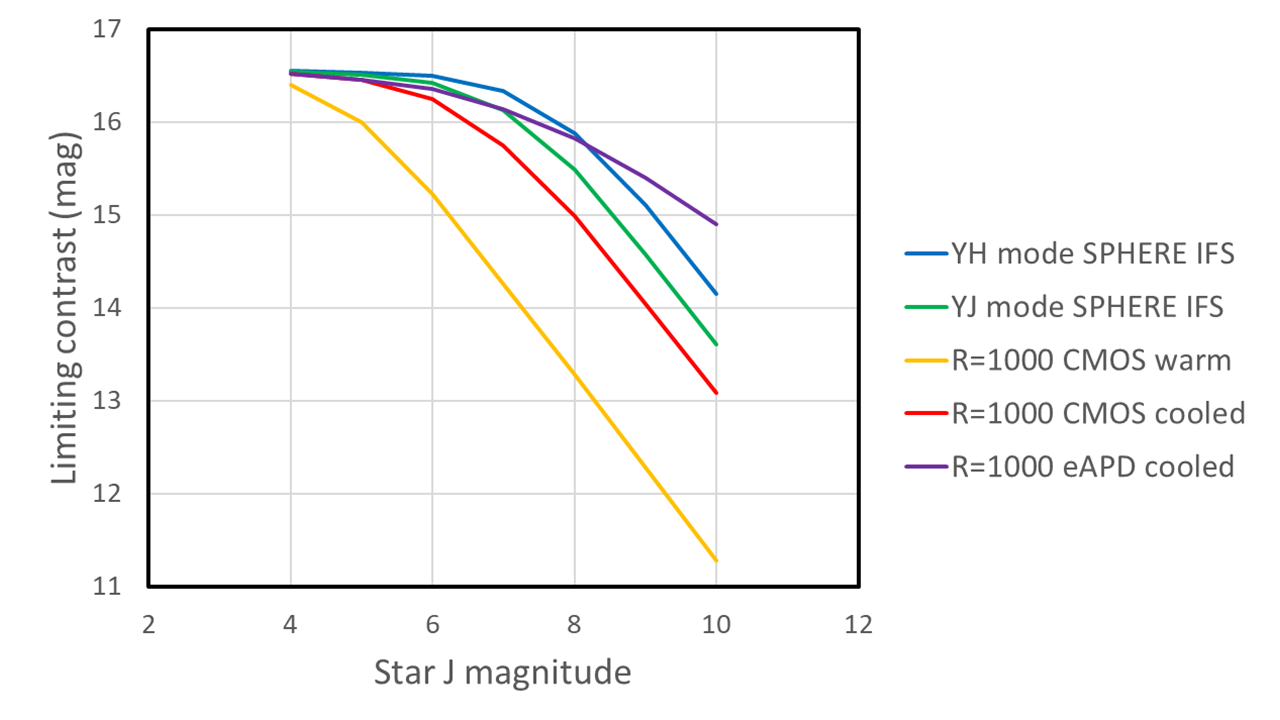}
\caption{Expected contrast as a function of the target J magnitude for the 2 modes in the SPHERE IFS and for a higher spectral resolution (R=1000) with various assumptions of temperature (warm vs cooled), and read out noise (CMOS vs. eAPD).}
\label{fig:IFScontrast}
\end{center}
\end{figure*}

The new spectro-imager, MEDRES, is intended to improve the atmosphere characterization of exoplanets  but also importantly to boost the detection by taking advantage of the cross correlation technique. When it comes to increasing the spectral resolution the noise budget does matter to achieve large contrasts. A model that includes the most important noise components has been developed for the SPHERE IFS and checked against real data. Extrapolating to larger spectral resolution, and plotting the contrast versus target star's magnitude clearly demonstrates the necessity of low noise detector and low thermal background in order to achieve contrast better than $10^6$ even for the faintest stars (Fig \ref{fig:IFScontrast}).  

The spectral resolution is a key parameter for MEDRES as it sets the performances in terms of detection and drives the spectral coverage and FoV. A first series of numerical simulations were performed to determine the optimal spectral resolution when using molecular mapping. Simulations are using Fourier-based formalism developed for the ZELDA experiment in SPHERE \cite{Vigan2019,Vigan2022}, which are representative of a bright star observation in moderate seeing conditions (0.7$''$ seeing, 5\,mas coherence time). The simulated wavefront maps include reconstructed tip and tilt, high order modes, as well as a model of aliasing and fitting errors. Beside, the coronagraph has the same features as the apodized Lyot coronagraph (ALC2 mask and STOP\_ALC2 diaphragm). Planets with L and T spectral types were injected at various angular separations from the inner working angle of the coronagraph (about 0.1$''$) out to the edge of the FoV. Then, we tested several spectral resolution ranging from 500 to 100000 in both the J and H bands. We found that for young giant planets the spectra of which are mostly H$_2$0 dominated, R$\approx$1000 provides the optimal signal to noise ratio. For other molecules like CO and CH$_4$, even lower spectral resolution of a few hundreds are sufficient.

\begin{figure*}[t]
\begin{center}
\includegraphics[trim = 0cm 0cm 0cm 0cm, width = 9.cm]{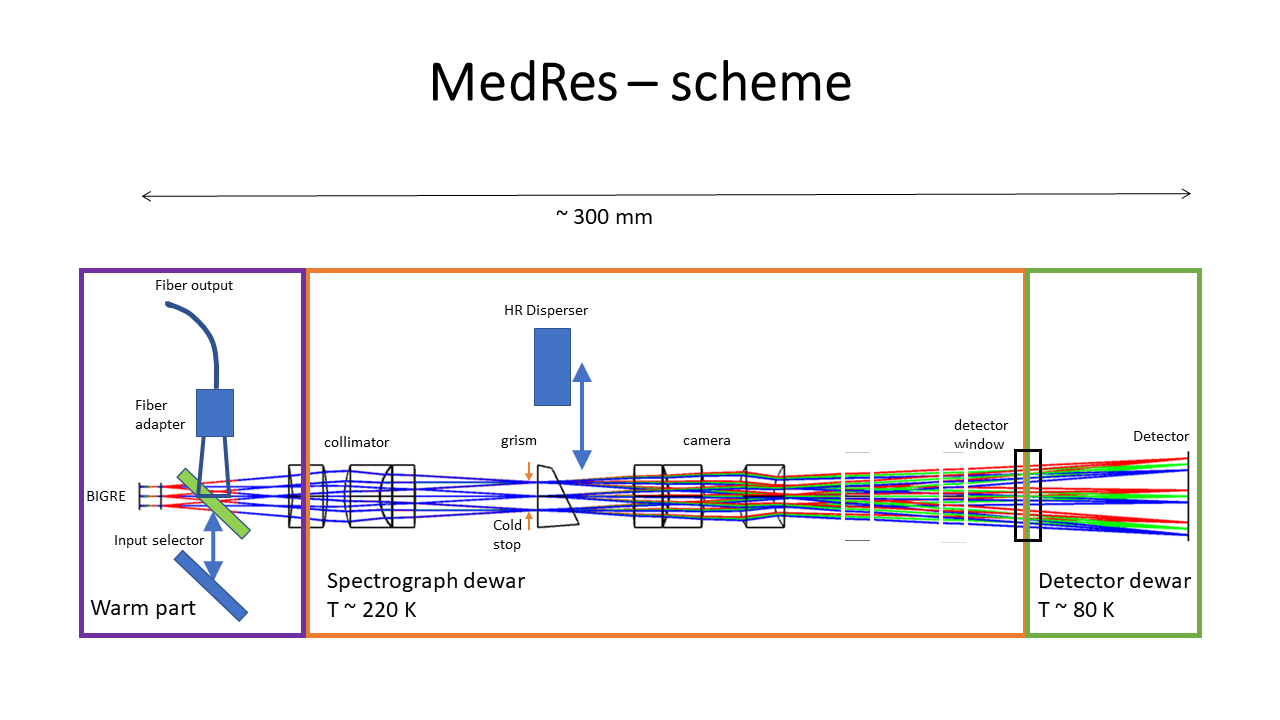}
\includegraphics[trim=0cm 5cm 0cm 2cm, clip, height=5.cm]{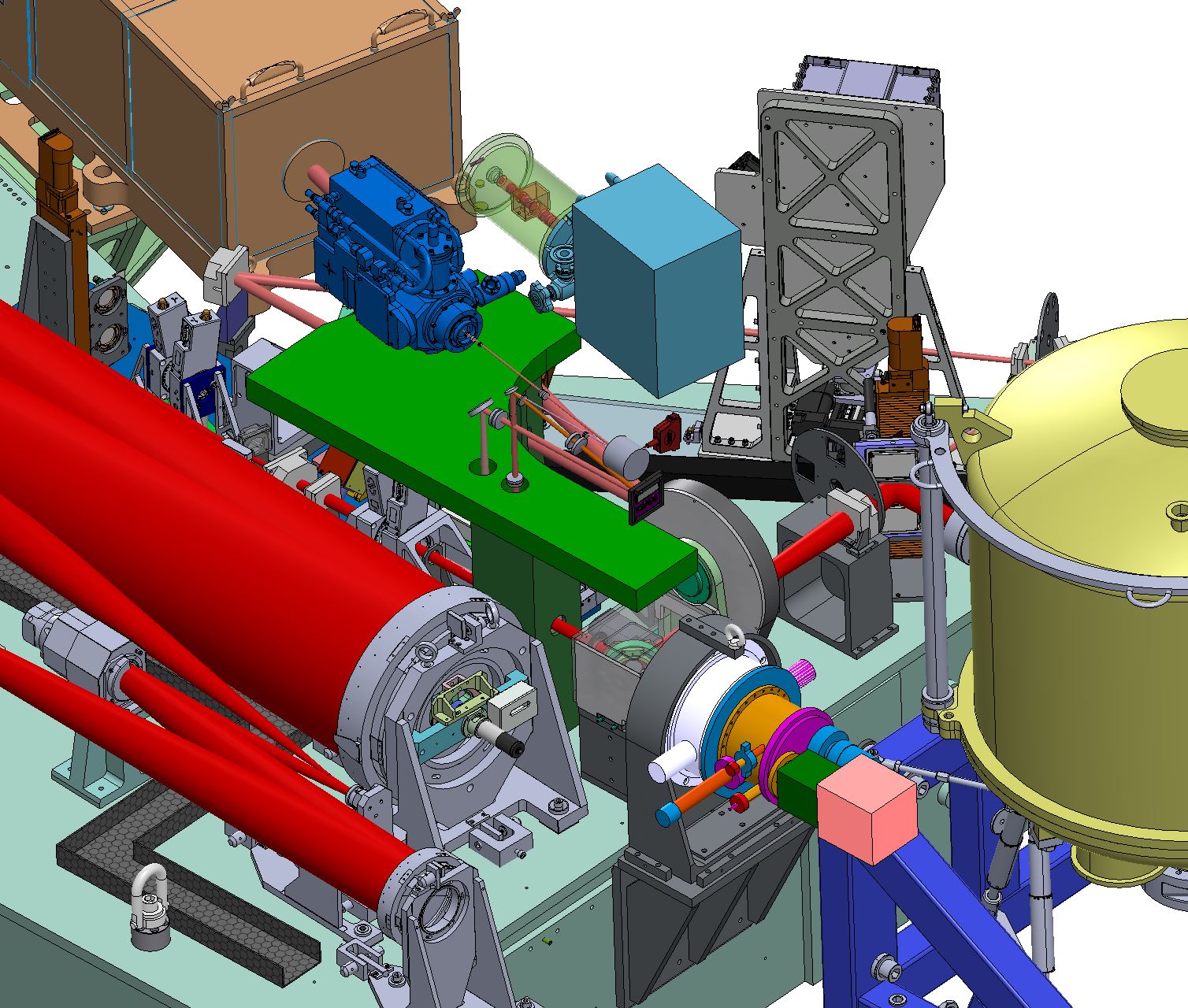}\\
\caption{Proposed optical design of MEDRES (left) showing the 3 different parts (IFU, spectrograph, detector), and the MEDRES implantation in SPHERE (right).}
\label{fig:medres}
\end{center}
\end{figure*}

The design of MEDRES is based on the same concept as the IFS, that is a BIGRE integral field unit configuration with two frames of microlenses \cite{Antichi2009}. With a spectral resolution of 1000 and a 1k\,$\times$\,1k detector, MEDRES achieves a 0.56$''\times 0.56''$ FoV in about 10\% of either the J or H band. To follow the aforementioned noise requirements, while the IFU itself is located in a warm part of the instrument, the spectrograph is cooled down to 220\,K and the detector is at 80\,K, the two latter being installed in two cryogenic vessels. We are considering an electron Avalanche Photodiode Array to achieve very low noise as again driven by the error budget. The whole assembly is about 30\,cm long. Similarly to SAXO+, MEDRES can be installed on top of the  SPHERE optical beam, 40\,cm above (Fig. \ref{fig:medres}). A full description of MEDRES is provided in Gratton et al. (this proceeding, paper 12184-164). To achieve higher spectral resolution (R=5000) which can be of interest for characterization purpose (at the expense of a lower detection level) we will also consider fiber fed IFU based on multi-core fibers with an exchange mechanism to switch between the 2 designs and a corresponding disperser inside the spectrograph (see Fig. \ref{fig:medres}). Finally, we plan also to explore a possible alternative design based on an image slicer (see Meyer et al., this proceeding, paper 12184-121).

\section{Conclusion}

The two new sub-systems, SAXO+ and MEDRES, which constitute the SPHERE upgrade are supported by the same consortium but follow two separate development plans.

SAXO+ is proposed as a technological demonstrator on-sky in the context of the Planetary Camera and Spectrograph (PCS) roadmap for the Extremely Large Telescope \cite{Kasper2021}. Indeed, an efficient extreme AO system is required for PCS to detect the light reflected by planets, possibly down to Super Earth size. However, designing this system for such a large telescope remains challenging, and upscaling of existing systems will not be sufficient. As of today, the solution to obtain a very high contrast ($10^9-10^{10}$ at 0.1$''$) with PCS necessarily involves a two-stage AO operating at high frequency. 
In mid 2022, SAXO+ is entering a feasibility study expected to last 18 months in which we will consolidate the opto-mechanical design.
This will be driven by numerical simulations the purpose of which is to define the main characteristics and the operating modes of the whole AO system, both SAXO and SAXO+, and to estimate performance in terms of contrasts then translated into exoplanet yields. Procurement can start in early 2024 followed with Assembly Integration and test in Europe in 2025. The on-sky implementation is expected early 2026. 

MEDRES is developed in parallel to SAXO+ with the aim to provide an optimized spectro-imager for both detecting and characterizing young giant planets, as a sound complement to other existing similar type of instruments, which either are general purpose facilities, or designed to reach very high spectral resolution for characterization only. MEDRES will be proposed as a visitor instrument for SPHERE.


\bibliography{sphereplus_spie_12184_62.bib} 

\begin{thebibliography}{10}

\bibitem{Beuzit2019}
{Beuzit}, J.~L., {Vigan}, A., {Mouillet}, D., {Dohlen}, K., {Gratton}, R.,
  {Boccaletti}, A., {Sauvage}, J.~F., {Schmid}, H.~M., {Langlois}, M., {Petit},
  C., {Baruffolo}, A., {Feldt}, M., {Milli}, J., {Wahhaj}, Z., {Abe}, L.,
  {Anselmi}, U., {Antichi}, J., {Barette}, R., {Baudrand}, J., {Baudoz}, P.,
  {Bazzon}, A., {Bernardi}, P., {Blanchard}, P., {Brast}, R., {Bruno}, P.,
  {Buey}, T., {Carbillet}, M., {Carle}, M., {Cascone}, E., {Chapron}, F.,
  {Charton}, J., {Chauvin}, G., {Claudi}, R., {Costille}, A., {De Caprio}, V.,
  {de Boer}, J., {Delboulb{\'e}}, A., {Desidera}, S., {Dominik}, C., {Downing},
  M., {Dupuis}, O., {Fabron}, C., {Fantinel}, D., {Farisato}, G., {Feautrier},
  P., {Fedrigo}, E., {Fusco}, T., {Gigan}, P., {Ginski}, C., {Girard}, J.,
  {Giro}, E., {Gisler}, D., {Gluck}, L., {Gry}, C., {Henning}, T., {Hubin}, N.,
  {Hugot}, E., {Incorvaia}, S., {Jaquet}, M., {Kasper}, M., {Lagadec}, E.,
  {Lagrange}, A.~M., {Le Coroller}, H., {Le Mignant}, D., {Le Ruyet}, B.,
  {Lessio}, G., {Lizon}, J.~L., {Llored}, M., {Lundin}, L., {Madec}, F.,
  {Magnard}, Y., {Marteaud}, M., {Martinez}, P., {Maurel}, D., {M{\'e}nard},
  F., {Mesa}, D., {M{\"o}ller-Nilsson}, O., {Moulin}, T., {Moutou}, C.,
  {Orign{\'e}}, A., {Parisot}, J., {Pavlov}, A., {Perret}, D., {Pragt}, J.,
  {Puget}, P., {Rabou}, P., {Ramos}, J., {Reess}, J.~M., {Rigal}, F., {Rochat},
  S., {Roelfsema}, R., {Rousset}, G., {Roux}, A., {Saisse}, M., {Salasnich},
  B., {Santambrogio}, E., {Scuderi}, S., {Segransan}, D., {Sevin}, A.,
  {Siebenmorgen}, R., {Soenke}, C., {Stadler}, E., {Suarez}, M., {Tiph{\`e}ne},
  D., {Turatto}, M., {Udry}, S., {Vakili}, F., {Waters}, L.~B.~F.~M., {Weber},
  L., {Wildi}, F., {Zins}, G., and {Zurlo}, A., ``{SPHERE: the exoplanet imager
  for the Very Large Telescope},'' {\em Astronomy \& Astrophysics}~{\bf 631},
  A155 (Nov 2019).

\bibitem{Cheetham2018}
{Cheetham}, A., {S{\'e}gransan}, D., {Peretti}, S., {Delisle}, J.~B.,
  {Hagelberg}, J., {Beuzit}, J.~L., {Forveille}, T., {Marmier}, M., {Udry}, S.,
  and {Wildi}, F., ``{Direct imaging of an ultracool substellar companion to
  the exoplanet host star HD 4113 A},'' {\em Astronomy \& Astrophysics}~{\bf
  614},  A16 (Jun 2018).

\bibitem{Maire2016}
{Maire}, A.-L., {Bonnefoy}, M., {Ginski}, C., {Vigan}, A., {Messina}, S.,
  {Mesa}, D., {Galicher}, R., {Gratton}, R., {Desidera}, S., {Kopytova}, T.~G.,
  {Millward}, M., {Thalmann}, C., {Claudi}, R.~U., {Ehrenreich}, D., {Zurlo},
  A., {Chauvin}, G., {Antichi}, J., {Baruffolo}, A., {Bazzon}, A., {Beuzit},
  J.-L., {Blanchard}, P., {Boccaletti}, A., {de Boer}, J., {Carle}, M.,
  {Cascone}, E., {Costille}, A., {De Caprio}, V., {Delboulb{\'e}}, A.,
  {Dohlen}, K., {Dominik}, C., {Feldt}, M., {Fusco}, T., {Girard}, J.~H.,
  {Giro}, E., {Gisler}, D., {Gluck}, L., {Gry}, C., {Henning}, T., {Hubin}, N.,
  {Hugot}, E., {Jaquet}, M., {Kasper}, M., {Lagrange}, A.-M., {Langlois}, M.,
  {Le Mignant}, D., {Llored}, M., {Madec}, F., {Martinez}, P., {Mawet}, D.,
  {Milli}, J., {M{\"o}ller-Nilsson}, O., {Mouillet}, D., {Moulin}, T.,
  {Moutou}, C., {Orign{\'e}}, A., {Pavlov}, A., {Petit}, C., {Pragt}, J.,
  {Puget}, P., {Ramos}, J., {Rochat}, S., {Roelfsema}, R., {Salasnich}, B.,
  {Sauvage}, J.-F., {Schmid}, H.~M., {Turatto}, M., {Udry}, S., {Vakili}, F.,
  {Wahhaj}, Z., {Weber}, L., and {Wildi}, F., ``{First light of the VLT planet
  finder SPHERE. II. The physical properties and the architecture of the young
  systems PZ Telescopii and HD 1160 revisited},'' {\em Astronomy \&
  Astrophysics}~{\bf 587},  A56 (Mar. 2016).

\bibitem{Delorme2017}
{Delorme}, P., {Schmidt}, T., {Bonnefoy}, M., {Desidera}, S., {Ginski}, C.,
  {Charnay}, B., {Lazzoni}, C., {Christiaens}, V., {Messina}, S., {D'Orazi},
  V., {Milli}, J., {Schlieder}, J.~E., {Gratton}, R., {Rodet}, L., {Lagrange},
  A.-M., {Absil}, O., {Vigan}, A., {Galicher}, R., {Hagelberg}, J., {Bonavita},
  M., {Lavie}, B., {Zurlo}, A., {Olofsson}, J., {Boccaletti}, A.,
  {Cantalloube}, F., {Mouillet}, D., {Chauvin}, G., {Hambsch}, F.-J.,
  {Langlois}, M., {Udry}, S., {Henning}, T., {Beuzit}, J.-L., {Mordasini}, C.,
  {Lucas}, P., {Marocco}, F., {Biller}, B., {Carson}, J., {Cheetham}, A.,
  {Covino}, E., {De Caprio}, V., {Delboulbe}, A., {Feldt}, M., {Girard}, J.,
  {Hubin}, N., {Maire}, A.-L., {Pavlov}, A., {Petit}, C., {Rouan}, D.,
  {Roelfsema}, R., and {Wildi}, F., ``{In-depth study of moderately young but
  extremely red, very dusty substellar companion HD 206893B},'' {\em Astronomy
  \& Astrophysics}~{\bf 608},  A79 (Dec. 2017).

\bibitem{Zurlo2016}
{Zurlo}, A., {Vigan}, A., {Galicher}, R., {Maire}, A.-L., {Mesa}, D.,
  {Gratton}, R., {Chauvin}, G., {Kasper}, M., {Moutou}, C., {Bonnefoy}, M.,
  {Desidera}, S., {Abe}, L., {Apai}, D., {Baruffolo}, A., {Baudoz}, P.,
  {Baudrand}, J., {Beuzit}, J.-L., {Blancard}, P., {Boccaletti}, A.,
  {Cantalloube}, F., {Carle}, M., {Cascone}, E., {Charton}, J., {Claudi},
  R.~U., {Costille}, A., {de Caprio}, V., {Dohlen}, K., {Dominik}, C.,
  {Fantinel}, D., {Feautrier}, P., {Feldt}, M., {Fusco}, T., {Gigan}, P.,
  {Girard}, J.~H., {Gisler}, D., {Gluck}, L., {Gry}, C., {Henning}, T.,
  {Hugot}, E., {Janson}, M., {Jaquet}, M., {Lagrange}, A.-M., {Langlois}, M.,
  {Llored}, M., {Madec}, F., {Magnard}, Y., {Martinez}, P., {Maurel}, D.,
  {Mawet}, D., {Meyer}, M.~R., {Milli}, J., {Moeller-Nilsson}, O., {Mouillet},
  D., {Orign{\'e}}, A., {Pavlov}, A., {Petit}, C., {Puget}, P., {Quanz}, S.~P.,
  {Rabou}, P., {Ramos}, J., {Rousset}, G., {Roux}, A., {Salasnich}, B.,
  {Salter}, G., {Sauvage}, J.-F., {Schmid}, H.~M., {Soenke}, C., {Stadler}, E.,
  {Suarez}, M., {Turatto}, M., {Udry}, S., {Vakili}, F., {Wahhaj}, Z., {Wildi},
  F., and {Antichi}, J., ``{First light of the VLT planet finder SPHERE. III.
  New spectrophotometry and astrometry of the HR 8799 exoplanetary system},''
  {\em Astronomy \& Astrophysics}~{\bf 587},  A57 (Mar. 2016).

\bibitem{Bonnefoy2016}
{Bonnefoy}, M., {Zurlo}, A., {Baudino}, J.~L., {Lucas}, P., {Mesa}, D.,
  {Maire}, A.-L., {Vigan}, A., {Galicher}, R., {Homeier}, D., {Marocco}, F.,
  {Gratton}, R., {Chauvin}, G., {Allard}, F., {Desidera}, S., {Kasper}, M.,
  {Moutou}, C., {Lagrange}, A.-M., {Antichi}, J., {Baruffolo}, A., {Baudrand},
  J., {Beuzit}, J.-L., {Boccaletti}, A., {Cantalloube}, F., {Carbillet}, M.,
  {Charton}, J., {Claudi}, R.~U., {Costille}, A., {Dohlen}, K., {Dominik}, C.,
  {Fantinel}, D., {Feautrier}, P., {Feldt}, M., {Fusco}, T., {Gigan}, P.,
  {Girard}, J.~H., {Gluck}, L., {Gry}, C., {Henning}, T., {Janson}, M.,
  {Langlois}, M., {Madec}, F., {Magnard}, Y., {Maurel}, D., {Mawet}, D.,
  {Meyer}, M.~R., {Milli}, J., {Moeller-Nilsson}, O., {Mouillet}, D., {Pavlov},
  A., {Perret}, D., {Pujet}, P., {Quanz}, S.~P., {Rochat}, S., {Rousset}, G.,
  {Roux}, A., {Salasnich}, B., {Salter}, G., {Sauvage}, J.-F., {Schmid}, H.~M.,
  {Sevin}, A., {Soenke}, C., {Stadler}, E., {Turatto}, M., {Udry}, S.,
  {Vakili}, F., {Wahhaj}, Z., and {Wildi}, F., ``{First light of the VLT planet
  finder SPHERE. IV. Physical and chemical properties of the planets around
  HR8799},'' {\em Astronomy \& Astrophysics}~{\bf 587},  A58 (Mar. 2016).

\bibitem{Chauvin2018}
{Chauvin}, G., {Gratton}, R., {Bonnefoy}, M., {Lagrange}, A.-M., {de Boer}, J.,
  {Vigan}, A., {Beust}, H., {Lazzoni}, C., {Boccaletti}, A., {Galicher}, R.,
  {Desidera}, S., {Delorme}, P., {Keppler}, M., {Lannier}, J., {Maire}, A.-L.,
  {Mesa}, D., {Meunier}, N., {Kral}, Q., {Henning}, T., {Menard}, F., {Moor},
  A., {Avenhaus}, H., {Bazzon}, A., {Janson}, M., {Beuzit}, J.-L., {Bhowmik},
  T., {Bonavita}, M., {Borgniet}, S., {Brandner}, W., {Cheetham}, A., {Cudel},
  M., {Feldt}, M., {Fontanive}, C., {Ginski}, C., {Hagelberg}, J.,
  {Janin-Potiron}, P., {Lagadec}, E., {Langlois}, M., {Le Coroller}, H.,
  {Messina}, S., {Meyer}, M., {Mouillet}, D., {Peretti}, S., {Perrot}, C.,
  {Rodet}, L., {Samland}, M., {Sissa}, E., {Olofsson}, J., {Salter}, G.,
  {Schmidt}, T., {Zurlo}, A., {Milli}, J., {van Boekel}, R., {Quanz}, S.,
  {Feautrier}, P., {Le Mignant}, D., {Perret}, D., {Ramos}, J., and {Rochat},
  S., ``{Investigating the young solar system analog HD 95086. A combined HARPS
  and SPHERE exploration},'' {\em Astronomy \& Astrophysics}~{\bf 617},  A76
  (Sept. 2018).

\bibitem{Samland2017}
{Samland}, M., {Molli{\`e}re}, P., {Bonnefoy}, M., {Maire}, A.-L.,
  {Cantalloube}, F., {Cheetham}, A.~C., {Mesa}, D., {Gratton}, R., {Biller},
  B.~A., {Wahhaj}, Z., {Bouwman}, J., {Brandner}, W., {Melnick}, D., {Carson},
  J., {Janson}, M., {Henning}, T., {Homeier}, D., {Mordasini}, C., {Langlois},
  M., {Quanz}, S.~P., {van Boekel}, R., {Zurlo}, A., {Schlieder}, J.~E.,
  {Avenhaus}, H., {Beuzit}, J.-L., {Boccaletti}, A., {Bonavita}, M., {Chauvin},
  G., {Claudi}, R., {Cudel}, M., {Desidera}, S., {Feldt}, M., {Fusco}, T.,
  {Galicher}, R., {Kopytova}, T.~G., {Lagrange}, A.-M., {Le Coroller}, H.,
  {Martinez}, P., {Moeller-Nilsson}, O., {Mouillet}, D., {Mugnier}, L.~M.,
  {Perrot}, C., {Sevin}, A., {Sissa}, E., {Vigan}, A., and {Weber}, L.,
  ``{Spectral and atmospheric characterization of 51 Eridani b using
  VLT/SPHERE},'' {\em Astronomy \& Astrophysics}~{\bf 603},  A57 (July 2017).

\bibitem{Chauvin2017}
{Chauvin}, G., {Desidera}, S., {Lagrange}, A.-M., {Vigan}, A., {Gratton}, R.,
  {Langlois}, M., {Bonnefoy}, M., {Beuzit}, J.-L., {Feldt}, M., {Mouillet}, D.,
  {Meyer}, M., {Cheetham}, A., {Biller}, B., {Boccaletti}, A., {D'Orazi}, V.,
  {Galicher}, R., {Hagelberg}, J., {Maire}, A.-L., {Mesa}, D., {Olofsson}, J.,
  {Samland}, M., {Schmidt}, T.~O.~B., {Sissa}, E., {Bonavita}, M., {Charnay},
  B., {Cudel}, M., {Daemgen}, S., {Delorme}, P., {Janin-Potiron}, P., {Janson},
  M., {Keppler}, M., {Le Coroller}, H., {Ligi}, R., {Marleau}, G.~D.,
  {Messina}, S., {Molli{\`e}re}, P., {Mordasini}, C., {M{\"u}ller}, A.,
  {Peretti}, S., {Perrot}, C., {Rodet}, L., {Rouan}, D., {Zurlo}, A.,
  {Dominik}, C., {Henning}, T., {Menard}, F., {Schmid}, H.-M., {Turatto}, M.,
  {Udry}, S., {Vakili}, F., {Abe}, L., {Antichi}, J., {Baruffolo}, A.,
  {Baudoz}, P., {Baudrand}, J., {Blanchard}, P., {Bazzon}, A., {Buey}, T.,
  {Carbillet}, M., {Carle}, M., {Charton}, J., {Cascone}, E., {Claudi}, R.,
  {Costille}, A., {Deboulbe}, A., {De Caprio}, V., {Dohlen}, K., {Fantinel},
  D., {Feautrier}, P., {Fusco}, T., {Gigan}, P., {Giro}, E., {Gisler}, D.,
  {Gluck}, L., {Hubin}, N., {Hugot}, E., {Jaquet}, M., {Kasper}, M., {Madec},
  F., {Magnard}, Y., {Martinez}, P., {Maurel}, D., {Le Mignant}, D.,
  {M{\"o}ller-Nilsson}, O., {Llored}, M., {Moulin}, T., {Orign{\'e}}, A.,
  {Pavlov}, A., {Perret}, D., {Petit}, C., {Pragt}, J., {Puget}, P., {Rabou},
  P., {Ramos}, J., {Rigal}, R., {Rochat}, S., {Roelfsema}, R., {Rousset}, G.,
  {Roux}, A., {Salasnich}, B., {Sauvage}, J.-F., {Sevin}, A., {Soenke}, C.,
  {Stadler}, E., {Suarez}, M., {Weber}, L., {Wildi}, F., {Antoniucci}, S.,
  {Augereau}, J.-C., {Baudino}, J.-L., {Brandner}, W., {Engler}, N., {Girard},
  J., {Gry}, C., {Kral}, Q., {Kopytova}, T., {Lagadec}, E., {Milli}, J.,
  {Moutou}, C., {Schlieder}, J., {Szul{\'a}gyi}, J., {Thalmann}, C., and
  {Wahhaj}, Z., ``{Discovery of a warm, dusty giant planet around HIP 65426},''
  {\em Astronomy \& Astrophysics}~{\bf 605},  L9 (Sept. 2017).

\bibitem{Keppler2018}
{Keppler}, M., {Benisty}, M., {M{\"u}ller}, A., {Henning}, T., {van Boekel},
  R., {Cantalloube}, F., {Ginski}, C., {van Holstein}, R.~G., {Maire}, A.-L.,
  {Pohl}, A., {Samland}, M., {Avenhaus}, H., {Baudino}, J.-L., {Boccaletti},
  A., {de Boer}, J., {Bonnefoy}, M., {Chauvin}, G., {Desidera}, S., {Langlois},
  M., {Lazzoni}, C., {Marleau}, G.-D., {Mordasini}, C., {Pawellek}, N.,
  {Stolker}, T., {Vigan}, A., {Zurlo}, A., {Birnstiel}, T., {Brandner}, W.,
  {Feldt}, M., {Flock}, M., {Girard}, J., {Gratton}, R., {Hagelberg}, J.,
  {Isella}, A., {Janson}, M., {Juhasz}, A., {Kemmer}, J., {Kral}, Q.,
  {Lagrange}, A.-M., {Launhardt}, R., {Matter}, A., {M{\'e}nard}, F., {Milli},
  J., {Molli{\`e}re}, P., {Olofsson}, J., {P{\'e}rez}, L., {Pinilla}, P.,
  {Pinte}, C., {Quanz}, S.~P., {Schmidt}, T., {Udry}, S., {Wahhaj}, Z.,
  {Williams}, J.~P., {Buenzli}, E., {Cudel}, M., {Dominik}, C., {Galicher}, R.,
  {Kasper}, M., {Lannier}, J., {Mesa}, D., {Mouillet}, D., {Peretti}, S.,
  {Perrot}, C., {Salter}, G., {Sissa}, E., {Wildi}, F., {Abe}, L., {Antichi},
  J., {Augereau}, J.-C., {Baruffolo}, A., {Baudoz}, P., {Bazzon}, A., {Beuzit},
  J.-L., {Blanchard}, P., {Brems}, S.~S., {Buey}, T., {De Caprio}, V.,
  {Carbillet}, M., {Carle}, M., {Cascone}, E., {Cheetham}, A., {Claudi}, R.,
  {Costille}, A., {Delboulb{\'e}}, A., {Dohlen}, K., {Fantinel}, D.,
  {Feautrier}, P., {Fusco}, T., {Giro}, E., {Gluck}, L., {Gry}, C., {Hubin},
  N., {Hugot}, E., {Jaquet}, M., {Le Mignant}, D., {Llored}, M., {Madec}, F.,
  {Magnard}, Y., {Martinez}, P., {Maurel}, D., {Meyer}, M.,
  {M{\"o}ller-Nilsson}, O., {Moulin}, T., {Mugnier}, L., {Orign{\'e}}, A.,
  {Pavlov}, A., {Perret}, D., {Petit}, C., {Pragt}, J., {Puget}, P., {Rabou},
  P., {Ramos}, J., {Rigal}, F., {Rochat}, S., {Roelfsema}, R., {Rousset}, G.,
  {Roux}, A., {Salasnich}, B., {Sauvage}, J.-F., {Sevin}, A., {Soenke}, C.,
  {Stadler}, E., {Suarez}, M., {Turatto}, M., and {Weber}, L., ``{Discovery of
  a planetary-mass companion within the gap of the transition disk around PDS
  70},'' {\em Astronomy \& Astrophysics}~{\bf 617},  A44 (Sept. 2018).

\bibitem{Muller2018}
{M{\"u}ller}, A., {Keppler}, M., {Henning}, T., {Samland}, M., {Chauvin}, G.,
  {Beust}, H., {Maire}, A.-L., {Molaverdikhani}, K., {van Boekel}, R.,
  {Benisty}, M., {Boccaletti}, A., {Bonnefoy}, M., {Cantalloube}, F.,
  {Charnay}, B., {Baudino}, J.-L., {Gennaro}, M., {Long}, Z.~C., {Cheetham},
  A., {Desidera}, S., {Feldt}, M., {Fusco}, T., {Girard}, J., {Gratton}, R.,
  {Hagelberg}, J., {Janson}, M., {Lagrange}, A.-M., {Langlois}, M., {Lazzoni},
  C., {Ligi}, R., {M{\'e}nard}, F., {Mesa}, D., {Meyer}, M., {Molli{\`e}re},
  P., {Mordasini}, C., {Moulin}, T., {Pavlov}, A., {Pawellek}, N., {Quanz},
  S.~P., {Ramos}, J., {Rouan}, D., {Sissa}, E., {Stadler}, E., {Vigan}, A.,
  {Wahhaj}, Z., {Weber}, L., and {Zurlo}, A., ``{Orbital and atmospheric
  characterization of the planet within the gap of the PDS 70 transition
  disk},'' {\em Astronomy \& Astrophysics}~{\bf 617},  L2 (Sept. 2018).

\bibitem{Lagrange2019}
{Lagrange}, A.-M., {Boccaletti}, A., {Langlois}, M., {Chauvin}, G., {Gratton},
  R., {Beust}, H., {Desidera}, S., {Milli}, J., {Bonnefoy}, M., {Cheetham}, A.,
  {Feldt}, M., {Meyer}, M., {Vigan}, A., {Biller}, B., {Bonavita}, M.,
  {Baudino}, J.-L., {Cantalloube}, F., {Cudel}, M., {Daemgen}, S., {Delorme},
  P., {D'Orazi}, V., {Girard}, J., {Fontanive}, C., {Hagelberg}, J., {Janson},
  M., {Keppler}, M., {Koypitova}, T., {Galicher}, R., {Lannier}, J., {Le
  Coroller}, H., {Ligi}, R., {Maire}, A.-L., {Mesa}, D., {Messina}, S.,
  {M{\"u}eller}, A., {Peretti}, S., {Perrot}, C., {Rouan}, D., {Salter}, G.,
  {Samland}, M., {Schmidt}, T., {Sissa}, E., {Zurlo}, A., {Beuzit}, J.-L.,
  {Mouillet}, D., {Dominik}, C., {Henning}, T., {Lagadec}, E., {M{\'e}nard},
  F., {Schmid}, H.-M., {Turatto}, M., {Udry}, S., {Bohn}, A.~J., {Charnay}, B.,
  {Gomez Gonzales}, C.~A., {Gry}, C., {Kenworthy}, M., {Kral}, Q., {Mordasini},
  C., {Moutou}, C., {van der Plas}, G., {Schlieder}, J.~E., {Abe}, L.,
  {Antichi}, J., {Baruffolo}, A., {Baudoz}, P., {Baudrand}, J., {Blanchard},
  P., {Bazzon}, A., {Buey}, T., {Carbillet}, M., {Carle}, M., {Charton}, J.,
  {Cascone}, E., {Claudi}, R., {Costille}, A., {Deboulbe}, A., {De Caprio}, V.,
  {Dohlen}, K., {Fantinel}, D., {Feautrier}, P., {Fusco}, T., {Gigan}, P.,
  {Giro}, E., {Gisler}, D., {Gluck}, L., {Hubin}, N., {Hugot}, E., {Jaquet},
  M., {Kasper}, M., {Madec}, F., {Magnard}, Y., {Martinez}, P., {Maurel}, D.,
  {Le Mignant}, D., {M{\"o}ller-Nilsson}, O., {Llored}, M., {Moulin}, T.,
  {Orign{\'e}}, A., {Pavlov}, A., {Perret}, D., {Petit}, C., {Pragt}, J.,
  {Szulagyi}, J., and {Wildi}, F., ``{Post-conjunction detection of {$\beta$}
  Pictoris b with VLT/SPHERE},'' {\em Astronomy \& Astrophysics}~{\bf 621},  L8
  (Jan. 2019).

\bibitem{Maire2019}
{Maire}, A.~L., {Rodet}, L., {Cantalloube}, F., {Galicher}, R., {Brandner}, W.,
  {Messina}, S., {Lazzoni}, C., {Mesa}, D., {Melnick}, D., {Carson}, J.,
  {Samland}, M., {Biller}, B.~A., {Boccaletti}, A., {Wahhaj}, Z., {Beust}, H.,
  {Bonnefoy}, M., {Chauvin}, G., {Desidera}, S., {Langlois}, M., {Henning}, T.,
  {Janson}, M., {Olofsson}, J., {Rouan}, D., {M{\'e}nard}, F., {Lagrange},
  A.~M., {Gratton}, R., {Vigan}, A., {Meyer}, M.~R., {Cheetham}, A., {Beuzit},
  J.~L., {Dohlen}, K., {Avenhaus}, H., {Bonavita}, M., {Claudi}, R., {Cudel},
  M., {Daemgen}, S., {D'Orazi}, V., {Fontanive}, C., {Hagelberg}, J., {Le
  Coroller}, H., {Perrot}, C., {Rickman}, E., {Schmidt}, T., {Sissa}, E.,
  {Udry}, S., {Zurlo}, A., {Abe}, L., {Orign{\'e}}, A., {Rigal}, F., {Rousset},
  G., {Roux}, A., and {Weber}, L., ``{Hint of curvature in the orbital motion
  of the exoplanet 51 Eridani b using 3 yr of VLT/SPHERE monitoring},'' {\em
  Astronomy \& Astrophysics}~{\bf 624},  A118 (Apr. 2019).

\bibitem{Benisty2015}
{Benisty}, M., {Juhasz}, A., {Boccaletti}, A., {Avenhaus}, H., {Milli}, J.,
  {Thalmann}, C., {Dominik}, C., {Pinilla}, P., {Buenzli}, E., {Pohl}, A.,
  {Beuzit}, J.-L., {Birnstiel}, T., {de Boer}, J., {Bonnefoy}, M., {Chauvin},
  G., {Christiaens}, V., {Garufi}, A., {Grady}, C., {Henning}, T., {Huelamo},
  N., {Isella}, A., {Langlois}, M., {M{\'e}nard}, F., {Mouillet}, D.,
  {Olofsson}, J., {Pantin}, E., {Pinte}, C., and {Pueyo}, L., ``{Asymmetric
  features in the protoplanetary disk MWC 758},'' {\em Astronomy \&
  Astrophysics}~{\bf 578},  L6 (June 2015).

\bibitem{vanBoekel2017}
{van Boekel}, R., {Henning}, T., {Menu}, J., {de Boer}, J., {Langlois}, M.,
  {M{\"u}ller}, A., {Avenhaus}, H., {Boccaletti}, A., {Schmid}, H.~M.,
  {Thalmann}, C., {Benisty}, M., {Dominik}, C., {Ginski}, C., {Girard}, J.~H.,
  {Gisler}, D., {Lobo Gomes}, A., {Menard}, F., {Min}, M., {Pavlov}, A.,
  {Pohl}, A., {Quanz}, S.~P., {Rabou}, P., {Roelfsema}, R., {Sauvage}, J.-F.,
  {Teague}, R., {Wildi}, F., and {Zurlo}, A., ``{Three Radial Gaps in the Disk
  of TW Hydrae Imaged with SPHERE},'' {\em Astrophysical Journal}~{\bf 837},
  132 (Mar. 2017).

\bibitem{Stolker2016}
{Stolker}, T., {Dominik}, C., {Avenhaus}, H., {Min}, M., {de Boer}, J.,
  {Ginski}, C., {Schmid}, H.~M., {Juhasz}, A., {Bazzon}, A., {Waters},
  L.~B.~F.~M., {Garufi}, A., {Augereau}, J.-C., {Benisty}, M., {Boccaletti},
  A., {Henning}, T., {Langlois}, M., {Maire}, A.-L., {M{\'e}nard}, F., {Meyer},
  M.~R., {Pinte}, C., {Quanz}, S.~P., {Thalmann}, C., {Beuzit}, J.-L.,
  {Carbillet}, M., {Costille}, A., {Dohlen}, K., {Feldt}, M., {Gisler}, D.,
  {Mouillet}, D., {Pavlov}, A., {Perret}, D., {Petit}, C., {Pragt}, J.,
  {Rochat}, S., {Roelfsema}, R., {Salasnich}, B., {Soenke}, C., and {Wildi},
  F., ``{Shadows cast on the transition disk of HD 135344B. Multiwavelength
  VLT/SPHERE polarimetric differential imaging},'' {\em Astronomy \&
  Astrophysics}~{\bf 595},  A113 (Nov. 2016).

\bibitem{Pohl2017}
{Pohl}, A., {Benisty}, M., {Pinilla}, P., {Ginski}, C., {de Boer}, J.,
  {Avenhaus}, H., {Henning}, T., {Zurlo}, A., {Boccaletti}, A., {Augereau},
  J.-C., {Birnstiel}, T., {Dominik}, C., {Facchini}, S., {Fedele}, D.,
  {Janson}, M., {Keppler}, M., {Kral}, Q., {Langlois}, M., {Ligi}, R., {Maire},
  A.-L., {M{\'e}nard}, F., {Meyer}, M., {Pinte}, C., {Quanz}, S.~P., {Sauvage},
  J.-F., {Sezestre}, {\'E}., {Stolker}, T., {Szul{\'a}gyi}, J., {van Boekel},
  R., {van der Plas}, G., {Villenave}, M., {Baruffolo}, A., {Baudoz}, P., {Le
  Mignant}, D., {Maurel}, D., {Ramos}, J., and {Weber}, L., ``{The
  Circumstellar Disk HD 169142: Gas, Dust, and Planets Acting in Concert?},''
  {\em Astrophysical Journal}~{\bf 850},  52 (Nov. 2017).

\bibitem{Ligi2018}
{Ligi}, R., {Vigan}, A., {Gratton}, R., {de Boer}, J., {Benisty}, M.,
  {Boccaletti}, A., {Quanz}, S.~P., {Meyer}, M., {Ginski}, C., {Sissa}, E.,
  {Gry}, C., {Henning}, T., {Beuzit}, J.-L., {Biller}, B., {Bonnefoy}, M.,
  {Chauvin}, G., {Cheetham}, A.~C., {Cudel}, M., {Delorme}, P., {Desidera}, S.,
  {Feldt}, M., {Galicher}, R., {Girard}, J., {Janson}, M., {Kasper}, M.,
  {Kopytova}, T., {Lagrange}, A.-M., {Langlois}, M., {Lecoroller}, H., {Maire},
  A.-L., {M{\'e}nard}, F., {Mesa}, D., {Peretti}, S., {Perrot}, C., {Pinilla},
  P., {Pohl}, A., {Rouan}, D., {Stolker}, T., {Samland}, M., {Wahhaj}, Z.,
  {Wildi}, F., {Zurlo}, A., {Buey}, T., {Fantinel}, D., {Fusco}, T., {Jaquet},
  M., {Moulin}, T., {Ramos}, J., {Suarez}, M., and {Weber}, L.,
  ``{Investigation of the inner structures around HD 169142 with VLT/SPHERE},''
  {\em Mont. Not. Roy. Ast. Soc.}~{\bf 473},  1774--1783 (Jan. 2018).

\bibitem{Lagrange2016}
{Lagrange}, A.-M., {Langlois}, M., {Gratton}, R., {Maire}, A.-L., {Milli}, J.,
  {Olofsson}, J., {Vigan}, A., {Bailey}, V., {Mesa}, D., {Chauvin}, G.,
  {Boccaletti}, A., {Galicher}, R., {Girard}, J.~H., {Bonnefoy}, M., {Samland},
  M., {Menard}, F., {Henning}, T., {Kenworthy}, M., {Thalmann}, C., {Beust},
  H., {Beuzit}, J.-L., {Brandner}, W., {Buenzli}, E., {Cheetham}, A., {Janson},
  M., {le Coroller}, H., {Lannier}, J., {Mouillet}, D., {Peretti}, S.,
  {Perrot}, C., {Salter}, G., {Sissa}, E., {Wahhaj}, Z., {Abe}, L., {Desidera},
  S., {Feldt}, M., {Madec}, F., {Perret}, D., {Petit}, C., {Rabou}, P.,
  {Soenke}, C., and {Weber}, L., ``{A narrow, edge-on disk resolved around HD
  106906 with SPHERE},'' {\em Astronomy \& Astrophysics}~{\bf 586},  L8 (Feb.
  2016).

\bibitem{Milli2017}
{Milli}, J., {Vigan}, A., {Mouillet}, D., {Lagrange}, A.-M., {Augereau}, J.-C.,
  {Pinte}, C., {Mawet}, D., {Schmid}, H.~M., {Boccaletti}, A., {Matr{\`a}}, L.,
  {Kral}, Q., {Ertel}, S., {Chauvin}, G., {Bazzon}, A., {M{\'e}nard}, F.,
  {Beuzit}, J.-L., {Thalmann}, C., {Dominik}, C., {Feldt}, M., {Henning}, T.,
  {Min}, M., {Girard}, J.~H., {Galicher}, R., {Bonnefoy}, M., {Fusco}, T., {de
  Boer}, J., {Janson}, M., {Maire}, A.-L., {Mesa}, D., {Schlieder}, J.~E., and
  {SPHERE Consortium}, ``{Near-infrared scattered light properties of the HR
  4796 A dust ring. A measured scattering phase function from 13.6 deg to 166.6
  deg},'' {\em Astronomy \& Astrophysics}~{\bf 599},  A108 (Mar. 2017).

\bibitem{Perrot2016}
{Perrot}, C., {Boccaletti}, A., {Pantin}, E., {Augereau}, J.-C., {Lagrange},
  A.-M., {Galicher}, R., {Maire}, A.-L., {Mazoyer}, J., {Milli}, J., {Rousset},
  G., {Gratton}, R., {Bonnefoy}, M., {Brandner}, W., {Buenzli}, E., {Langlois},
  M., {Lannier}, J., {Mesa}, D., {Peretti}, S., {Salter}, G., {Sissa}, E.,
  {Chauvin}, G., {Desidera}, S., {Feldt}, M., {Vigan}, A., {Di Folco}, E.,
  {Dutrey}, A., {P{\'e}ricaud}, J., {Baudoz}, P., {Benisty}, M., {De Boer}, J.,
  {Garufi}, A., {Girard}, J.~H., {Menard}, F., {Olofsson}, J., {Quanz}, S.~P.,
  {Mouillet}, D., {Christiaens}, V., {Casassus}, S., {Beuzit}, J.-L.,
  {Blanchard}, P., {Carle}, M., {Fusco}, T., {Giro}, E., {Hubin}, N., {Maurel},
  D., {Moeller-Nilsson}, O., {Sevin}, A., and {Weber}, L., ``{Discovery of
  concentric broken rings at sub-arcsec separations in the HD 141569A gas-rich,
  debris disk with VLT/SPHERE},'' {\em Astronomy \& Astrophysics}~{\bf 590},
  L7 (May 2016).

\bibitem{Feldt2017}
{Feldt}, M., {Olofsson}, J., {Boccaletti}, A., {Maire}, A.~L., {Milli}, J.,
  {Vigan}, A., {Langlois}, M., {Henning}, T., {Moor}, A., {Bonnefoy}, M.,
  {Wahhaj}, Z., {Desidera}, S., {Gratton}, R., {K{\'o}sp{\'a}l}, {\'A}.,
  {Abraham}, P., {Menard}, F., {Chauvin}, G., {Lagrange}, A.~M., {Mesa}, D.,
  {Salter}, G., {Buenzli}, E., {Lannier}, J., {Perrot}, C., {Peretti}, S., and
  {Sissa}, E., ``{SPHERE/SHINE reveals concentric rings in the debris disk of
  HIP 73145},'' {\em Astronomy \& Astrophysics}~{\bf 601},  A7 (May 2017).

\bibitem{Boccaletti2015}
{Boccaletti}, A., {Thalmann}, C., {Lagrange}, A.-M., {Janson}, M., {Augereau},
  J.-C., {Schneider}, G., {Milli}, J., {Grady}, C., {Debes}, J., {Langlois},
  M., {Mouillet}, D., {Henning}, T., {Dominik}, C., {Maire}, A.-L., {Beuzit},
  J.-L., {Carson}, J., {Dohlen}, K., {Engler}, N., {Feldt}, M., {Fusco}, T.,
  {Ginski}, C., {Girard}, J.~H., {Hines}, D., {Kasper}, M., {Mawet}, D.,
  {M{\'e}nard}, F., {Meyer}, M.~R., {Moutou}, C., {Olofsson}, J., {Rodigas},
  T., {Sauvage}, J.-F., {Schlieder}, J., {Schmid}, H.~M., {Turatto}, M.,
  {Udry}, S., {Vakili}, F., {Vigan}, A., {Wahhaj}, Z., and {Wisniewski}, J.,
  ``{Fast-moving features in the debris disk around AU Microscopii},'' {\em
  Nature}~{\bf 526},  230--232 (Oct. 2015).

\bibitem{Vigan2021}
{Vigan}, A., {Fontanive}, C., {Meyer}, M., {Biller}, B., {Bonavita}, M.,
  {Feldt}, M., {Desidera}, S., {Marleau}, G.~D., {Emsenhuber}, A., {Galicher},
  R., {Rice}, K., {Forgan}, D., {Mordasini}, C., {Gratton}, R., {Le Coroller},
  H., {Maire}, A.~L., {Cantalloube}, F., {Chauvin}, G., {Cheetham}, A.,
  {Hagelberg}, J., {Lagrange}, A.~M., {Langlois}, M., {Bonnefoy}, M., {Beuzit},
  J.~L., {Boccaletti}, A., {D'Orazi}, V., {Delorme}, P., {Dominik}, C.,
  {Henning}, T., {Janson}, M., {Lagadec}, E., {Lazzoni}, C., {Ligi}, R.,
  {Menard}, F., {Mesa}, D., {Messina}, S., {Moutou}, C., {M{\"u}ller}, A.,
  {Perrot}, C., {Samland}, M., {Schmid}, H.~M., {Schmidt}, T., {Sissa}, E.,
  {Turatto}, M., {Udry}, S., {Zurlo}, A., {Abe}, L., {Antichi}, J.,
  {Asensio-Torres}, R., {Baruffolo}, A., {Baudoz}, P., {Baudrand}, J.,
  {Bazzon}, A., {Blanchard}, P., {Bohn}, A.~J., {Brown Sevilla}, S.,
  {Carbillet}, M., {Carle}, M., {Cascone}, E., {Charton}, J., {Claudi}, R.,
  {Costille}, A., {De Caprio}, V., {Delboulb{\'e}}, A., {Dohlen}, K., {Engler},
  N., {Fantinel}, D., {Feautrier}, P., {Fusco}, T., {Gigan}, P., {Girard},
  J.~H., {Giro}, E., {Gisler}, D., {Gluck}, L., {Gry}, C., {Hubin}, N.,
  {Hugot}, E., {Jaquet}, M., {Kasper}, M., {Le Mignant}, D., {Llored}, M.,
  {Madec}, F., {Magnard}, Y., {Martinez}, P., {Maurel}, D.,
  {M{\"o}ller-Nilsson}, O., {Mouillet}, D., {Moulin}, T., {Orign{\'e}}, A.,
  {Pavlov}, A., {Perret}, D., {Petit}, C., {Pragt}, J., {Puget}, P., {Rabou},
  P., {Ramos}, J., {Rickman}, E.~L., {Rigal}, F., {Rochat}, S., {Roelfsema},
  R., {Rousset}, G., {Roux}, A., {Salasnich}, B., {Sauvage}, J.~F., {Sevin},
  A., {Soenke}, C., {Stadler}, E., {Suarez}, M., {Wahhaj}, Z., {Weber}, L., and
  {Wildi}, F., ``{The SPHERE infrared survey for exoplanets (SHINE). III. The
  demographics of young giant exoplanets below 300 au with SPHERE},'' {\em
  Astronomy \& Astrophysics}~{\bf 651},  A72 (July 2021).

\bibitem{Fernandes2019}
{Fernandes}, R.~B., {Mulders}, G.~D., {Pascucci}, I., {Mordasini}, C., and
  {Emsenhuber}, A., ``{Hints for a Turnover at the Snow Line in the Giant
  Planet Occurrence Rate},'' {\em Astrophysical Journal}~{\bf 874},  81 (Mar.
  2019).

\bibitem{Langlois2021}
{Langlois}, M., {Gratton}, R., {Lagrange}, A.~M., {Delorme}, P., {Boccaletti},
  A., {Bonnefoy}, M., {Maire}, A.~L., {Mesa}, D., {Chauvin}, G., {Desidera},
  S., {Vigan}, A., {Cheetham}, A., {Hagelberg}, J., {Feldt}, M., {Meyer}, M.,
  {Rubini}, P., {Le Coroller}, H., {Cantalloube}, F., {Biller}, B., {Bonavita},
  M., {Bhowmik}, T., {Brandner}, W., {Daemgen}, S., {D'Orazi}, V., {Flasseur},
  O., {Fontanive}, C., {Galicher}, R., {Girard}, J., {Janin-Potiron}, P.,
  {Janson}, M., {Keppler}, M., {Kopytova}, T., {Lagadec}, E., {Lannier}, J.,
  {Lazzoni}, C., {Ligi}, R., {Meunier}, N., {Perreti}, A., {Perrot}, C.,
  {Rodet}, L., {Romero}, C., {Rouan}, D., {Samland}, M., {Salter}, G., {Sissa},
  E., {Schmidt}, T., {Zurlo}, A., {Mouillet}, D., {Denis}, L., {Thi{\'e}baut},
  E., {Milli}, J., {Wahhaj}, Z., {Beuzit}, J.~L., {Dominik}, C., {Henning}, T.,
  {M{\'e}nard}, F., {M{\"u}ller}, A., {Schmid}, H.~M., {Turatto}, M., {Udry},
  S., {Abe}, L., {Antichi}, J., {Allard}, F., {Baruffolo}, A., {Baudoz}, P.,
  {Baudrand}, J., {Bazzon}, A., {Blanchard}, P., {Carbillet}, M., {Carle}, M.,
  {Cascone}, E., {Charton}, J., {Claudi}, R., {Costille}, A., {De Caprio}, V.,
  {Delboulb{\'e}}, A., {Dohlen}, K., {Fantinel}, D., {Feautrier}, P., {Fusco},
  T., {Gigan}, P., {Giro}, E., {Gisler}, D., {Gluck}, L., {Gry}, C., {Hubin},
  N., {Hugot}, E., {Jaquet}, M., {Kasper}, M., {Le Mignant}, D., {Llored}, M.,
  {Madec}, F., {Magnard}, Y., {Martinez}, P., {Maurel}, D., {Messina}, S.,
  {M{\"o}ller-Nilsson}, O., {Mugnier}, L., {Moulin}, T., {Orign{\'e}}, A.,
  {Pavlov}, A., {Perret}, D., {Petit}, C., {Pragt}, J., {Puget}, P., {Rabou},
  P., {Ramos}, J., {Rigal}, F., {Rochat}, S., {Roelfsema}, R., {Rousset}, G.,
  {Roux}, A., {Salasnich}, B., {Sauvage}, J.~F., {Sevin}, A., {Soenke}, C.,
  {Stadler}, E., {Suarez}, M., {Weber}, L., {Wildi}, F., and {Rickman}, E.,
  ``{The SPHERE infrared survey for exoplanets (SHINE). II. Observations, data
  reduction and analysis, detection performances, and initial results},'' {\em
  Astronomy \& Astrophysics}~{\bf 651},  A71 (July 2021).

\bibitem{Boccaletti2020a}
{Boccaletti}, A., {Di Folco}, E., {Pantin}, E., {Dutrey}, A., {Guilloteau}, S.,
  {Tang}, Y.~W., {Pi{\'e}tu}, V., {Habart}, E., {Milli}, J., {Beck}, T.~L., and
  {Maire}, A.~L., ``{Possible evidence of ongoing planet formation in AB
  Aurigae. A showcase of the SPHERE/ALMA synergy},'' {\em Astronomy \&
  Astrophysics}~{\bf 637},  L5 (May 2020).

\bibitem{Benisty2021}
{Benisty}, M., {Bae}, J., {Facchini}, S., {Keppler}, M., {Teague}, R.,
  {Isella}, A., {Kurtovic}, N.~T., {P{\'e}rez}, L.~M., {Sierra}, A., {Andrews},
  S.~M., {Carpenter}, J., {Czekala}, I., {Dominik}, C., {Henning}, T.,
  {Menard}, F., {Pinilla}, P., and {Zurlo}, A., ``{A Circumplanetary Disk
  around PDS70c},'' {\em Astrophysical Journal Letter}~{\bf 916},  L2 (July
  2021).

\bibitem{Fusco2014}
{Fusco}, T., {Sauvage}, J.~F., {Petit}, C., {Costille}, A., {Dohlen}, K.,
  {Mouillet}, D., {Beuzit}, J.~L., {Kasper}, M., {Suarez}, M., {Soenke}, C.,
  {Fedrigo}, E., {Downing}, M., {Baudoz}, P., {Sevin}, A., {Perret}, D.,
  {Barufolo}, A., {Salasnich}, B., {Puget}, P., {Feautrier}, F., {Rochat}, S.,
  {Moulin}, T., {Deboulb{\'e}}, A., {Hugot}, E., {Vigan}, A., {Mawet}, D.,
  {Girard}, J., and {Hubin}, N., ``{Final performance and lesson-learned of
  SAXO, the VLT-SPHERE extreme AO: from early design to on-sky results},'' in
  [{\em Adaptive Optics Systems IV}{\nolinebreak\hspace{0.1em}]},  {Marchetti},
  E., {Close}, L.~M., and {Vran}, J.-P., eds., {\em Society of Photo-Optical
  Instrumentation Engineers (SPIE) Conference Series} {\bf 9148},  91481U (Aug.
  2014).

\bibitem{Sauvage2016}
{Sauvage}, J.-F., {Fusco}, T., {Petit}, C., {Costille}, A., {Mouillet}, D.,
  {Beuzit}, J.-L., {Dohlen}, K., {Kasper}, M., {Suarez}, M., {Soenke}, C.,
  {Baruffolo}, A., {Salasnich}, B., {Rochat}, S., {Fedrigo}, E., {Baudoz}, P.,
  {Hugot}, E., {Sevin}, A., {Perret}, D., {Wildi}, F., {Downing}, M.,
  {Feautrier}, P., {Puget}, P., {Vigan}, A., {O'Neal}, J., {Girard}, J.,
  {Mawet}, D., {Schmid}, H.~M., and {Roelfsema}, R., ``{SAXO: the extreme
  adaptive optics system of SPHERE (I) system overview and global laboratory
  performance},'' {\em Journal of Astronomical Telescopes, Instruments, and
  Systems}~{\bf 2},  025003 (Apr. 2016).

\bibitem{Vigan2018}
{Vigan}, A., {Otten}, G.~P.~P.~L., {Muslimov}, E., {Dohlen}, K., {Philipps},
  M.~W., {Seemann}, U., {Beuzit}, J.~L., {Dorn}, R., {Kasper}, M., {Mouillet},
  D., {Baraffe}, I., and {Reiners}, A., ``{Bringing high-spectral resolution to
  VLT/SPHERE with a fiber coupling to VLT/CRIRES+},'' in [{\em Ground-based and
  Airborne Instrumentation for Astronomy VII}{\nolinebreak\hspace{0.1em}]},
  {\em in Proc. of SPIE} {\bf 10702},  1070236 (Jul 2018).

\bibitem{Otten2021}
{Otten}, G.~P.~P.~L., {Vigan}, A., {Muslimov}, E., {N'Diaye}, M., {Choquet},
  E., {Seemann}, U., {Dohlen}, K., {Houll{\'e}}, M., {Cristofari}, P.,
  {Phillips}, M.~W., {Charles}, Y., {Baraffe}, I., {Beuzit}, J.~L., {Costille},
  A., {Dorn}, R., {El Morsy}, M., {Kasper}, M., {Lopez}, M., {Mordasini}, C.,
  {Pourcelot}, R., {Reiners}, A., and {Sauvage}, J.~F., ``{Direct
  characterization of young giant exoplanets at high spectral resolution by
  coupling SPHERE and CRIRES+},'' {\em Astronomy \& Astrophysics}~{\bf 646},
  A150 (Feb. 2021).

\bibitem{Vigan2022spie}
{Vigan}, A., {Lopez}, M., {El Morsy}, M., {Muslimov}, E., {Viret}, A., {Zins},
  G., {Murray}, G., {Costille}, A., {Otten}, G.~P.~P.~L., {Seemann}, U.,
  {Anwand-Heerwart}, H., {Dohlen}, K., {Blanchard}, P., {Garcia}, J.,
  {Charles}, Y., {Tchoubaklian}, N., {Ely}, T., {Phillips}, M., {Paufique}, J.,
  {Beuzit}, J.~L., {Houll{\'e}}, M., {Costes}, J., {Pourcelot}, R., {Baraffe},
  I., {Dorn}, R., {Jaquet}, M., {Kasper}, M., {Reiners}, A., {Smette}, A.,
  {Blanco}, L., {Pallanca}, L., {Carlotti}, A., {Choquet}, {\'E}., {Mouillet},
  D., and {N'Diaye}, M., ``{Connecting SPHERE and CRIRES+ for the
  characterisation of young exoplanets at high spectral resolution: status
  update of VLT/HiRISE},'' {\em arXiv e-prints} ,  arXiv:2207.06436 (July
  2022).

\bibitem{Ferreira2018}
{Ferreira}, F., {Gratadour}, D., {Sevin}, A., {Doucet}, N., {Vidal}, F., {Deo},
  V., and {Gendron}, E., ``{Real-time end-to-end AO simulations at ELT scale on
  multiple GPUs with the COMPASS platform},'' in [{\em Adaptive Optics Systems
  VI}{\nolinebreak\hspace{0.1em}]},  {Close}, L.~M., {Schreiber}, L., and
  {Schmidt}, D., eds., {\em Society of Photo-Optical Instrumentation Engineers
  (SPIE) Conference Series} {\bf 10703},  1070347 (July 2018).

\bibitem{Carbillet2011}
Carbillet, M., Bendjoya, P., Abe, L., Guerri, G., Boccaletti, A., Daban, J.-B.,
  Dohlen, K., Ferrari, A., Robbe-Dubois, S., Douet, R., and Vakili, F.,
  ``{Apodized Lyot coronagraph for SPHERE/VLT. I. Detailed numerical study},''
  {\em Experimental Astronomy}~{\bf 30},  39 -- 58 (05 2011).

\bibitem{Vigan2019}
Vigan, A., N'Diaye, M., Dohlen, K., Sauvage, J.-F., Milli, J., Zins, G., Petit,
  C., Wahhaj, Z., Cantalloube, F., Caillat, A., Costille, A., Merrer, J.~L.,
  Carlotti, A., Beuzit, J.~L., and Mouillet, D., ``{Calibration of quasi-static
  aberrations in exoplanet direct-imaging instruments with a Zernike phase-mask
  sensor},'' {\em Astronomy \& Astrophysics}~{\bf 629},  A11 -- 16 (08 2019).

\bibitem{Potier2020}
Potier, A., Galicher, R., Baudoz, P., Huby, E., Milli, J., Wahhaj, Z.,
  Boccaletti, A., Vigan, A., N’Diaye, M., and Sauvage, J.-F., ``{Increasing
  the raw contrast of VLT/SPHERE with the dark hole technique},'' {\em
  Astronomy \& Astrophysics}~{\bf 638},  A117 (06 2020).

\bibitem{Hoeijmakers2018}
Hoeijmakers, H.~J., Schwarz, H., Snellen, I. A.~G., Kok, R. J.~d., Bonnefoy,
  M., Chauvin, G., Lagrange, A.~M., and Girard, J.~H., ``{Medium-resolution
  integral-field spectroscopy for high-contrast exoplanet imaging},'' {\em
  Astronomy \& Astrophysics}~{\bf 617},  A144 -- 11 (10 2018).

\bibitem{Petrus2020}
Petrus, S., Bonnefoy, M., Chauvin, G., Charnay, B., Marleau, G.-D., Gratton,
  R., Lagrange, A.-M., Rameau, J., Mordasini, C., Nowak, M., Delorme, P.,
  Boccaletti, A., Carlotti, A., Houllé, M., Vigan, A., Allard, F., Desidera,
  S., D’Orazi, V., Hoeijmakers, H.~J., Wyttenbach, A., and Lavie, B.,
  ``{Medium-resolution spectrum of the exoplanet HIP 65426 b},'' {\em Astronomy
  \& Astrophysics}~{\bf 648},  A59 (12 2020).

\bibitem{Vigan2022}
{Vigan}, A., {Dohlen}, K., {N'Diaye}, M., {Cantalloube}, F., {Girard}, J.~H.,
  {Milli}, J., {Sauvage}, J.~F., {Wahhaj}, Z., {Zins}, G., {Beuzit}, J.~L.,
  {Caillat}, A., {Costille}, A., {Le Merrer}, J., {Mouillet}, D., and
  {Tourenq}, S., ``{Calibration of quasi-static aberrations in exoplanet
  direct-imaging instruments with a Zernike phase-mask sensor. IV. Temporal
  stability of non-common path aberrations in VLT/SPHERE},'' {\em Astronomy \&
  Astrophysics}~{\bf 660},  A140 (Apr. 2022).

\bibitem{Antichi2009}
Antichi, J., Dohlen, K., Gratton, R.~G., Mesa, D., Claudi, R.~U., Giro, E.,
  Boccaletti, A., Mouillet, D., Puget, P., and Beuzit, J.-L., ``{BIGRE: A Low
  Cross-Talk Integral Field Unit Tailored for Extrasolar Planets Imaging
  Spectroscopy},'' {\em The Astrophysical Journal}~{\bf 695},  1042 -- 1057 (04
  2009).

\bibitem{Kasper2021}
{Kasper}, M., {Cerpa Urra}, N., {Pathak}, P., {Bonse}, M., {Nousiainen}, J.,
  {Engler}, B., {Heritier}, C.~T., {Kammerer}, J., {Leveratto}, S., {Rajani},
  C., {Bristow}, P., {Le Louarn}, M., {Madec}, P.~Y., {Str{\"o}bele}, S.,
  {Verinaud}, C., {Glauser}, A., {Quanz}, S.~P., {Helin}, T., {Keller}, C.,
  {Snik}, F., {Boccaletti}, A., {Chauvin}, G., {Mouillet}, D., {Kulcs{\'a}r},
  C., and {Raynaud}, H.~F., ``{PCS {\textemdash} A Roadmap for Exoearth Imaging
  with the ELT},'' {\em The Messenger}~{\bf 182},  38--43 (Mar. 2021).

\end{thebibliography}
\bibliographystyle{spiebib} 

\end{document}